\documentclass[a4paper,11pt]{article}

\usepackage{jheppub}
\usepackage{amsmath}
\usepackage{graphicx}
\usepackage{multirow}
\usepackage{bm}
\usepackage{bbm}
\usepackage{color}
\usepackage{slashed}
\usepackage{scalefnt}
\usepackage{grffile}
\allowdisplaybreaks

\title{Three-loop matching coefficients for  heavy flavor-changing    currents and the phenomenological applications}
\author[a]{Wei Tao}
\author[a]{Zhen-Jun Xiao\footnote{Corresponding author}}
\author[a,b]{Ruilin Zhu\footnote{Corresponding author}}
\affiliation[a]{Department of Physics and Institute of Theoretical Physics, Nanjing Normal University, Nanjing, Jiangsu 210023, China}
\affiliation[b]{ CAS Key Laboratory of Theoretical Physics and Peng Huanwu Innovation Research Center, Institute of Theoretical Physics,
	Chinese Academy of Sciences, Beijing 100190, China}

\emailAdd{taowei@njnu.edu.cn}
\emailAdd{xiaozhenjun@njnu.edu.cn}
\emailAdd{rlzhu@njnu.edu.cn}

\abstract{
	Within the framework of non-relativistic QCD (NRQCD) factorization,	we compute the matching coefficients between full Quantum Chromodynamics (QCD) and
	NRQCD  for the heavy flavor-changing    vector, axial-vector, scalar		and pseudo-scalar currents up to next-to-next-to-next-to-leading  order (N$^3$LO).
	We accomplish the analytical expressions for the three-loop  renormalization constants and the corresponding anomalous dimensions
	for all of	the four NRQCD  currents  with two different heavy flavors.
	The three-loop QCD corrections to the   matching coefficients turn out to be significantly
	larger than lower order corrections. By employing the scale relation, we obtain  the N$^3$LO corrections to the wave functions at the origin for the vector $B_c^*$ meson and the pseudo-scalar $B_c$ meson   from the known result 
	for the equal-mass  heavy quarkonium in potential NRQCD (pNRQCD). We find  large cancellations at the third order between    the   matching coefficients and the wave functions at the origin, and obtain the convergent  decay constants of  $B^*_{c}$ and $B_{c}$  up to N$^3$LO. We present the complete  perturbative QCD predictions for   the decay constants, leptonic decay widths, and  branching ratios    of the   beauty-charmed mesons.
	}

\keywords{Three-loop Calculation, NRQCD, Matching Coefficients, Convergent Decay Constants,   Beauty-charmed Mesons}
\preprint{~}

\begin{document}
	
\maketitle


\section{Introduction}

Double heavy quark systems such as the $J/\psi$, $\Upsilon$, $B_c$,  and $B_c^*$ mesons are an ideal probe to study  Quantum Chromodynamics (QCD).
The non-relativistic Quantum Chromodynamics  (NRQCD) effective theory  is a powerful theoretical framework to deal with
the production and decay of double heavy quark systems where  quarks move with small relative
velocity $u$~\cite{Bodwin:1994jh}.
Within the framework of NRQCD factorization, a physical observable can be separated into a perturbatively calculable short-distance coefficient (matching coefficient) multiplied with the nonperturbative  NRQCD long-distance matrix element (wave function at the origin).
Thus NRQCD   provides the possibility to  systematically  study double heavy quark systems with  higher order calculation  in   powers of  two small parameters  $\alpha_{s}$ and  $u$.

The     matching coefficients  can be calculated by matching between  perturbative QCD and  perturbative NRQCD.
In the past three decades, various  matching coefficients of double heavy quark systems have been computed to higher order using NRQCD effective theory.
For equal heavy quark   masses case, the  two-loop correction to the vector current was first obtained in Refs.~\cite{Beneke:1997jm,Czarnecki:1997vz}.
Then two-loop corrections to  vector, axial-vector, scalar and pseudo-scalar currents were accomplished  in
in Ref.~\cite{Kniehl:2006qw}.
The  three-loop correction to the vector current was investigated in  a variety of  literature ~\cite{Kniehl:2002yv,Marquard:2006qi,Marquard:2009bj,Beneke:2013jia,Marquard:2014pea,Feng:2022vvk}.
And three-loop calculation allowing for all four currents were available  in
 Refs.~\cite{Piclum:2007an,Egner:2021lxd,Egner:2022jot}.
The phenomenological applications for the leptonic
decays of  $J/\psi$, $\Upsilon$  and the threshold production of top quark-antiquark pairs
can be found in the  literature~\cite{Beneke:1997jm,Czarnecki:1997vz,Kniehl:2002yv,Piclum:2007an,Marquard:2006qi,Beneke:2013jia,Marquard:2014pea,Beneke:2015kwa,Beneke:2014qea,Egner:2021lxd,Feng:2022vvk}.
For two different heavy quark   masses case, the first
one-loop calculation for  pseudo-scalar current can be found in Ref.~\cite{Braaten:1995ej}.
And  in Refs.~\cite{Hwang:1999fc,Lee:2010ts} one can find the  one-loop  QCD corrections to
 pseudo-scalar and  vector     currents  combined with higher order relativistic corrections.
Two-loop corrections to  pseudo-scalar and  vector     currents were calculated in the literature ~\cite{Onishchenko:2003ui,Chen:2015csa,Tao:2022qxa}.
At the three-loop level, the  pseudo-scalar,  vector and  scalar     currents have been numerically evaluated in Ref.~\cite{Feng:2022ruy}, Ref.~\cite{Sang:2022tnh} and Ref.~\cite{Tao:2022hos}, respectively.
For phenomenological applications of matching coefficients in $B_c$  and $B_c^*$ mesons,  one can refer to various  literature~\cite{Hwang:1999fc,Onishchenko:2003ui,Chen:2015csa,Feng:2022ruy,Broadhurst:1994se,Kiselev:1998wb,Bell:2010mg,Tao:2022yur,Tao:2022qxa}.

The   matching coefficients and the wave function at the origin are two  important building blocks of phenomenological predictions for double heavy quark systems. 
For a complete N$^3$LO perturbative QCD calculation of physical  quantities such as decay constants, decay widths, cross sections, we need consider  the N$^3$LO corrections to both the   matching coefficients and the wave functions at the origin~\cite{Beneke:2015kwa,Beneke:2014qea,Egner:2021lxd}.
The higher-order perturbative correction to  the wave function at the origin can be calculated using the potential NRQCD effective theory~\cite{Pineda:1997bj,Beneke:1999zr,Brambilla:1999xf,Luke:1999kz,Brambilla:2004jw}.
For the  heavy quarkonia with two equal masses such as $J/\psi$ and $\Upsilon$, the wave functions at the origin up to N$^3$LO  are  available  in various literature~\cite{Beneke:2007uf,Beneke:2013jia,Beneke:2014qea,Peset:2015vvi,Shen:2015cta,Schroder:1998vy,Schuller:2008rxa,Beneke:2005hg,Beneke:2007gj,Beneke:2007pj,Beneke:2016kkb}.
For the unequal mass case, the NLO result of the wave function at the
  origin can be obtained from Ref.~\cite{Penin:2004xi} and various research on the perturbative  corrections to  potentials can be found in Refs.~\cite{Anzai:2018eua,MorenoTorres:2019uxb,Peset:2015vvi,PesetMartin:2016lus,Pineda:2011aw}. Nevertheless, higher-order corrections to the wave function at the origin for unequal masses are still missing in the literature.

In this paper we will complete the computation of three-loop QCD corrections to the     matching coefficients  for    vector, axial-vector, scalar		and pseudo-scalar currents with two different heavy quark masses, where the axial-vector case is achieved for the first time. We will not attempt to calculate the rigorous N$^3$LO expression of the wave function at origin for the double heavy quark system with different masses based on the pNRQCD effective theory, but employ the scale relation to obtain the   N$^3$LO correction to the wave functions at the origin for the vector $B_c^*$ meson and the pseudo-scalar $B_c$ meson   from  the known result for the equal-mass  heavy quarkonium. After  the three-loop matching coefficients are combined with the N$^3$LO corrections to the wave functions at the origin, the resultant decay constants for $B_c^*$ and $B_c$   will test the perturbative convergence in the NRQCD effective theory.

The rest of the paper is organized as follows.
In Sec.~\ref{Matchingformula}, we introduce the  matching formula between QCD and NRQCD.
In Sec.~\ref{Calculationprocedure},  we describe details of our calculation procedure.
In Sec.~\ref{ZjNRQCD}, we  present the three-loop analytical results of  the  renormalization constants and corresponding anomalous dimensions  in NRQCD.
In Sec.~\ref{Matching coefficients}, we present our numeric results of the  matching coefficients up to N$^3$LO.
In Sec.~\ref{Phenomenological}, the phenomenological applications of three-loop matching coefficients to the $c\bar{b}$ mesons are presented.
Sec.~\ref{Summary} contains a summary.

\section{Matching formula ~\label{Matchingformula}}
The heavy flavor-changing  currents in the full QCD are defined by
\begin{align}\label{QCDcurrents}
&j_v^\mu = \bar{\psi}_b \gamma^\mu \psi_c,
\nonumber \\&
j_p     = \bar{\psi}_b {\rm i}\gamma_5 \psi_c,
\nonumber \\&
j_a^\mu = \bar{\psi}_b \gamma^\mu\gamma_5 \psi_c,
\nonumber \\&
j_s     = \bar{\psi}_b \psi_c,
\end{align}
which  can be expanded  in terms of NRQCD currents as follows,
\begin{align}\label{expandcurrents}
&j_v^0 = 0+ {\mathcal O}(u^{2}),
\nonumber \\&
j_v^i = \mathcal{C}_{v}\tilde{j}_v^i + {\mathcal O}(u^{2}),
\nonumber \\&
j_p     = \mathcal{C}_{p}\tilde{j}_p + {\mathcal O}(u^{2}),
\nonumber \\&
j_a^0 = \mathcal{C}_{a,0}\tilde{j}_a^0 + {\mathcal O}(u^{2}),
\nonumber \\&
j_a^i = \mathcal{C}_{a,i}\tilde{j}_a^i + {\mathcal O}(u^{3}),
\nonumber \\&
j_s     = \mathcal{C}_{s}\tilde{j}_s + {\mathcal O}(u^{3}),
\end{align}
where $u$  refers   to the small relative velocity   between the bottom and charm,
 $\mathcal{C}_{v},\mathcal{C}_{p},\mathcal{C}_{a,0},\mathcal{C}_{a,i}$, $\mathcal{C}_{s}$
 are the matching coefficients for the   heavy flavor-changing      vector, pseudo-scalar, the zeroth component of axial-vector, the space-like component of axial-vector, and scalar currents, respectively. And the NRQCD currents read
 \begin{align}\label{NRQCDcurrents}
&\tilde{j}_v^i = \varphi_b^\dagger \sigma^i \chi_c,
\nonumber \\&
\tilde{j}_p     = -{\rm i}\,\varphi_b^\dagger \chi_c,
\nonumber \\&
\tilde{j}_a^0 = \varphi_b^\dagger \chi_c,
\nonumber \\&
\tilde{j}_a^i = \frac{1}{4 m_r} \varphi_b^\dagger[\sigma^i,\vec{k}\cdot\vec{\sigma}] \chi_c,
\nonumber \\&
\tilde{j}_s     = -\frac{1}{2m_r} \varphi_b^\dagger \vec{k}\cdot\vec{\sigma}\chi_c,
\end{align}
where  $\varphi_b^\dagger$ and $\chi_c$ denote 2-component Pauli spinor  fields annihilating the $\bar{b}$ and $c$ quarks, respectively, and $\tilde{j}_p=-{\rm i}\,\tilde{j}_a^0$ means $\mathcal{C}_{p}=\mathcal{C}_{a,0}$.
$|\vec{k}|= m_r u$  refers   to the small half relative spatial  momentum  between the bottom and charm.   $m_r=m_b m_c/(m_b+m_c)$ is the reduced  mass with bottom mass $m_b$ and charm mass $m_c$.

The heavy flavor-changing  currents in Eq.~\eqref{QCDcurrents} and Eq.~\eqref{NRQCDcurrents} can be related to  the on-shell unrenormalized  vertex functions in QCD and NRQCD
which we denote by $\Gamma_{J}$ and $\widetilde{\Gamma}_{J}$ with
$J\in\{v,p,(a,0),(a,i),s\}$, respectively.
Then the matching coefficients can be determined through the conventional perturbative matching procedure. Namely, one  performs renormalization for  the on-shell vertex functions in both perturbative QCD and perturbative
NRQCD sides, then solves the matching coefficient  order by order in $\alpha_s$.
The matching formula with renormalization procedure reads
\begin{align} \label{matchformularenormalization}
\sqrt{Z_{2,b} Z_{2,c} } \,Z_{J}  \, \Gamma_{J} =
\mathcal{C}_{J}(\mu_f,\mu,m_b,m_c) \, \sqrt{\widetilde{Z}_{2,b} \widetilde{Z}_{2,c} } \,
{\widetilde Z}_{J}^{-1} \, \widetilde{\Gamma}_{J} + {
	\mathcal O}(u^2),
\end{align}
where  the left part in the equation represents the renormalization of full QCD current while the right part represents the renormalization of NRQCD current.
${\mathcal O}(u^2)$ denotes higher order relativistic corrections in powers of the  relative velocity $u$ between the bottom quark  $\bar{b}$ and the charm quark $c$,
and in this paper we will calculate 
higher-order QCD corrections up to ${\mathcal O}(\alpha_s^3)$   but at the lowest order in $u$.
 At the leading-order (LO), we set $\mathcal{C}_{J}=1$,
while in a fixed high order QCD correction calculation,
  the matching coefficient $\mathcal{C}_{J}(\mu_f,\mu,m_b,m_c)$ depends NRQCD factorization scale $\mu_f$ and QCD  renormalization scale $\mu$.
In on-shell ($\mathrm{OS}$) scheme,   NRQCD  quark field renormalization constants $\widetilde{Z}_{2,b}=\widetilde{Z}_{2,c}=1$.
 $\widetilde{Z}_{J}$ is NRQCD current renormalization constant in the modified-minimal-subtraction ($\mathrm{\overline{MS}}$) scheme and $\widetilde{Z}_{p}=\widetilde{Z}_{a,0}$.
 $Z_{J}$ is QCD  current on-shell  renormalization constant, i.e., $Z_v=Z_{a,0}=Z_{a,i}=1$, $Z_p=Z_s=(m_b Z_{m,b}+m_c Z_{m,c})/(m_b+m_c)$. $Z_{2}$ and $Z_{m}$ are QCD  quark field and mass on-shell renormalization constants, respectively.
 The three-loop analytical results of  the QCD quark field and mass on-shell  renormalization constants  allowing for two different non-zero quark masses  can be found in literature~\cite{Bekavac:2007tk,Marquard:2016dcn,Fael:2020bgs}, which can be  evaluated to high numerical precision  with the package {\texttt{PolyLogTools}}~\cite{Duhr:2019tlz}.
 The QCD coupling $\mathrm{\overline{MS}}$ renormalization constant can be found in literature~\cite{Mitov:2006xs,Chetyrkin:1997un,vanRitbergen:1997va}.

\section{Calculation procedure~\label{Calculationprocedure}}

Our high-order calculation  consists of the following steps.
\begin{itemize}
	\item
	First, we use {\texttt{FeynCalc}}~\cite{Shtabovenko:2020gxv}  to obtain Feynman diagrams and corresponding Feynman amplitudes. By {\texttt{\$Apart}}~\cite{Feng:2012iq}, we decompose  every Feyman amplitude  into several Feynman integral families.
	\item
	Second, we use {\texttt{FIRE}}~\cite{Smirnov:2019qkx}/{\texttt{Kira}}~\cite{Klappert:2020nbg}/{\texttt{FiniteFlow}}~\cite{Peraro:2019svx} based on Integration by Parts (IBP)~\cite{Chetyrkin:1981qh} to reduce every Feynman integral family to master integral family.
	\item
	Third,
	based on symmetry among different integral families and using  {\texttt{Kira}} + {\texttt{FIRE}} + {\texttt{Mathematica\,code}}, we can realize  integral reduction among different integral families, and further on, the reduction from all of master integral families to the minimal set~\cite{Fael:2020njb} of master  integral families.
	\item
	Last, we use {\texttt{AMFlow}}~\cite{Liu:2022chg}, which is a proof-of-concept implementation of the auxiliary mass flow method~\cite{Liu:2017jxz}, equipped with {\texttt{Kira}}~\cite{Klappert:2020nbg}/{\texttt{FiniteFlow}}~\cite{Peraro:2019svx} to calculate the minimal set  family by family.
\end{itemize}
In order to obtain the finite results of  high-order QCD corrections, one has to perform the  conventional renormalization procedure~\cite{Chen:2015csa,Kniehl:2006qw,Bonciani:2008wf,Davydychev:1997vh}. Equivalently, We can also use diagrammatic renormalization method~\cite{deOliveira:2022eeq}  with the aid of the package {\texttt{FeynCalc}}~\cite{Shtabovenko:2020gxv}, which at N$^3$LO sums contributions from three-loop diagrams and four  kinds of counter-term diagrams, i.e.,  tree diagram   inserted with one $\alpha_s^3$-order counter-term vertex, one-loop diagram inserted with one $\alpha_s^2$-order counter-term vertex, one-loop diagram inserted with two $\alpha_s$-order counter-term vertexes, two-loop diagrams inserted with one $\alpha_s$-order counter-term vertex. Our final finite results by these two renormalization methods are in agreement with each other.

\begin{figure}[thb]
	\center{
		\includegraphics*[scale=0.7]{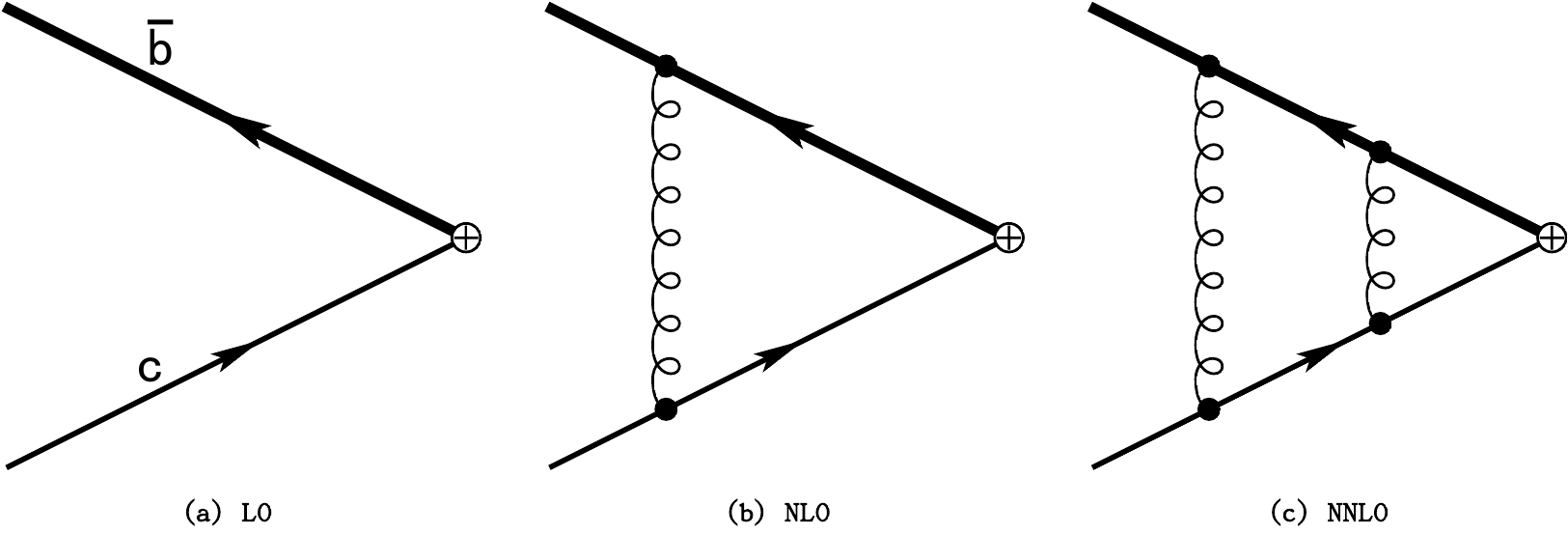}
		\caption {\label{fig:aiupto2looppics} Representative Feynman diagrams 	for the QCD vertex function with the heavy flavor-changing   currents  up to two-loop order.	 The cross ``$\bigoplus$'' implies the insertion of a certain heavy flavor-changing   current.	}}
\end{figure}

\begin{figure}[thb]
	\center{
		\includegraphics*[scale=0.9]{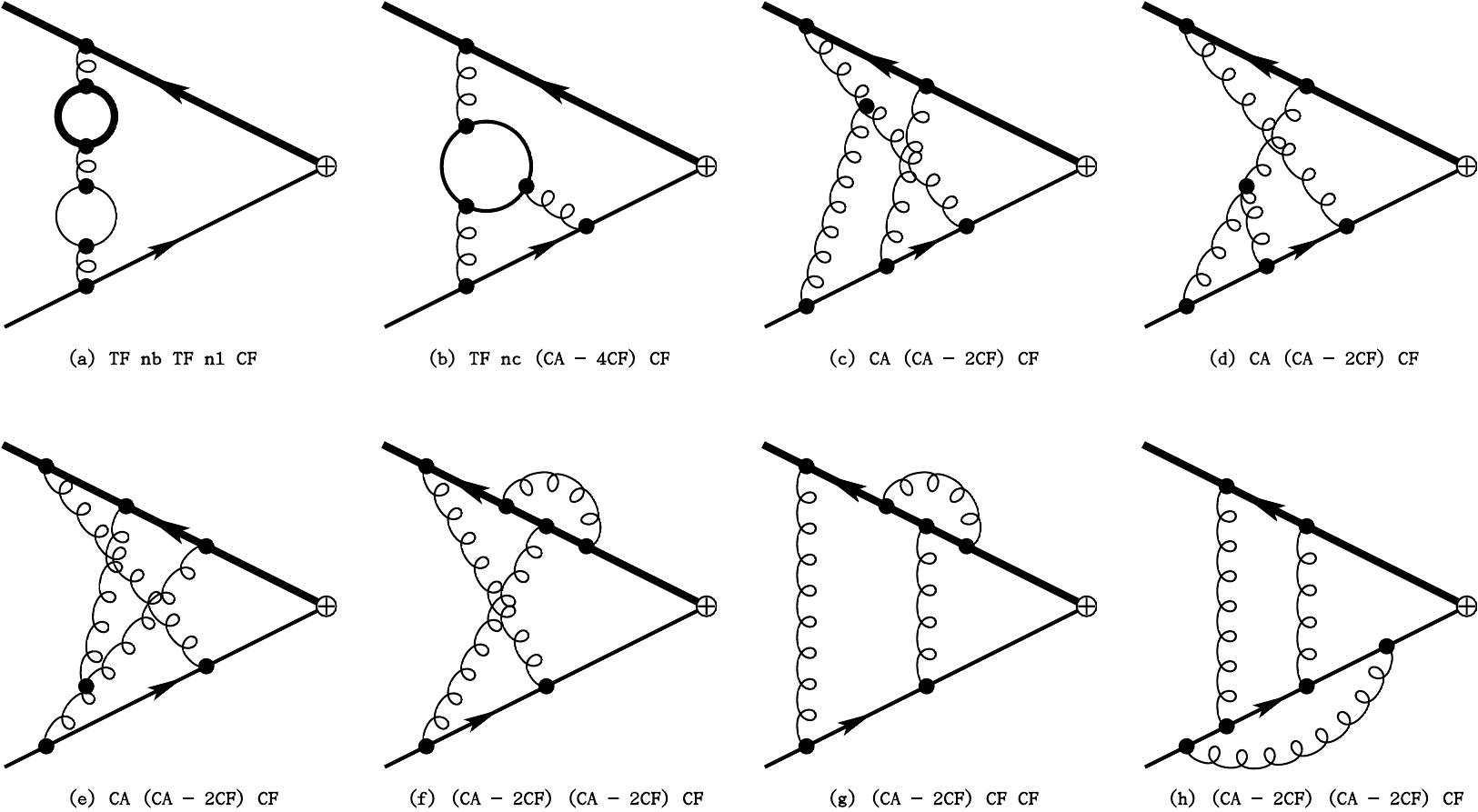}
		\caption {\label{fig:ai3looppics} Representative Feynman diagrams  labelled with corresponding color factors 	for the QCD vertex function with the heavy flavor-changing  currents   at three-loop order.	 The cross ``$\bigoplus$'' implies the insertion of a certain heavy flavor-changing   current.
			The thickest, thick, thinnest solid	closed circles represent the bottom loop with mass $m_b$, the charm loop with mass $m_c$, and the massless quark loop, respectively.}}
\end{figure}

We want to mention that all contributions up to NNLO have been evaluated for general gauge parameter $\xi$ and the NNLO results of the  matching coefficients   for the heavy flavor-changing   currents   are all independent of $\xi$, which constitutes an important check on our calculation.
At N$^3$LO, we work in  Feynman gauge. To check  the correctness of
our results, we also calculated the three-loop matching coefficient  for the  flavor-changing heavy quark pseudo-scalar current with keeping  the linear $\xi$-dependence terms and we have
verified these terms vanish  in the final results of the matching coefficient. Therefore, we conjecture the matching coefficients, the NRQCD current renormalization constants and the corresponding anomalous dimensions for flavor-changing heavy quark currents  are gauge invariant up to all order of $\alpha_s$.
By FeynCalc, there are 1, 1, 13, 268 bare Feynman diagrams for the QCD vertex function with every heavy flavor-changing   current  at tree, one-loop, two-loop, three-loop orders in $\alpha_s$, respectively.
Some representative Feynman diagrams up to  three loops  are displayed in Fig.~\ref{fig:aiupto2looppics} and Fig.~\ref{fig:ai3looppics}.
In the calculation of multi-loop diagrams, we have allowed for $n_b$ bottom quarks with mass $m_b$, $n_c$ charm quarks  with mass $m_c$ and $n_l$ massless quarks appearing in the quark loop.
To facilitate our calculation, we take full advantage of computing numerically. Namely, before generating amplitudes, $m_b$ and $m_c$ are chosen to be particular rational numbers values~\cite{Bronnum-Hansen:2021olh,Chen:2022vzo,Chen:2022mre}.

Following the literature~\cite{Kniehl:2006qw}, we employ the projectors constructed for the heavy flavor-changing   currents   to obtain intended QCD vertex functions, which means one need extend    the projectors for   various   currents   with equal heavy quark masses in Eq.(7)  and Eq.(8) of Ref.~\cite{Kniehl:2006qw} to the different heavy quark masses case. We choose $q_1=\frac{m_c}{m_b+m_c}q+k$ and $q_2=\frac{m_b}{m_b+m_c}q-k$ denoting the on-shell charm and bottom momentum, respectively, and present the projectors for the heavy flavor-changing  currents  as
\begin{align}
P_{(v),\mu} =& \frac{1}{2(D-1)(m_b+m_c)^2}
\left(\frac{m_c}{m_b+m_c}\slashed{q} + m_c
\right) \gamma_\mu \left(-\frac{m_b}{m_b+m_c}\slashed{q} + m_b
\right),
\\
P_{(p)} =& \frac{1}{2(m_b+m_c)^2}
\left(\frac{m_c}{m_b+m_c}\slashed{q} + m_c
\right) \gamma_5 \left(-\frac{m_b}{m_b+m_c}\slashed{q} + m_b
\right),
\\
P_{(a,0),\mu} =& -\frac{1}{2(m_b+m_c)^2}
\left(\frac{m_c}{m_b+m_c}\slashed{q} + m_c
\right) \gamma_\mu \gamma_5 \left(-\frac{m_b}{m_b+m_c}\slashed{q} + m_b
\right),
\\
P_{(a,i),\mu} =& -\frac{1}{2(m_b+m_c)^2} \bigg\{
\frac{1}{D-1}\frac{m_c}{m_b+m_c}\left(\frac{m_c}{m_b+m_c}\slashed{q} + m_c
\right) \gamma_\mu \gamma_5\left(\frac{m_b}{m_b+m_c}\slashed{q} + m_b
\right)
\nonumber\\&
+\frac{1}{D-1}\frac{m_b}{m_b+m_c}\left(-\frac{m_c}{m_b+m_c}\slashed{q} + m_c
\right) \gamma_\mu \gamma_5\left(-\frac{m_b}{m_b+m_c}\slashed{q} + m_b
\right)
\nonumber\\&
- \frac{1}{D-2}\frac{2 m_b m_c}{m_b+m_c} \left(\frac{m_c}{m_b+m_c}\slashed{q} + m_c
\right) \frac{-k_\mu + \gamma_\mu \slashed{k} }{k^2}
\gamma_5 \left(-\frac{m_b}{m_b+m_c}\slashed{q} + m_b\right) \bigg\},\label{projectorai}
\\
P_{(s)} = &\frac{1}{2(m_b+m_c)^2} \bigg\{ \frac{m_c}{m_b+m_c}\left(\frac{m_c}{m_b+m_c}\slashed{q} + m_c
\right) {\bf 1} \left(\frac{m_b}{m_b+m_c}\slashed{q} + m_b\right)
\nonumber\\&
+\frac{m_b}{m_b+m_c}\left(-\frac{m_c}{m_b+m_c}\slashed{q} + m_c
\right) {\bf 1} \left(-\frac{m_b}{m_b+m_c}\slashed{q} + m_b\right)
\nonumber\\&
+\frac{2 m_b m_c}{m_b+m_c}
\left(\frac{m_c}{m_b+m_c}\slashed{q} + m_c\right) \frac{\slashed{k}}{k^2}
\left(-\frac{m_b}{m_b+m_c}\slashed{q} + m_b\right) \bigg\},\label{projectors}
\end{align}
where the small momentum $k$ refers to  relative movement  between the bottom and charm, $q$ represents the total momentum of the bottom and charm, $q_1^2=m_c^2$, $q_2^2=m_b^2$, $q^2=(m_b+m_c)^2+\mathcal{O}{(k^2)}$~\cite{Zhu:2017lqu}, $q\cdot k=0$. With the help of projectors, the  on-shell heavy flavor-changing current vertex functions in full QCD can be obtained as
\begin{align}
&\Gamma_v = \mbox{Tr}\left[ P_{(v),\mu} \Gamma_{(v)}^{\mu} \right]\,,
\nonumber\\&
\Gamma_p = \mbox{Tr}\left[ P_{(p)} \Gamma_{(p)} \right]\,,
\nonumber\\&
\Gamma_{a,0} = \mbox{Tr}\left[ P_{(a,0),\mu} \Gamma_{(a)}^{\mu} \right]\,,
\nonumber\\&
\Gamma_{a,i} = \mbox{Tr}\left[ P_{(a,i),\mu} \Gamma_{(a)}^{\mu} \right]\,,
\nonumber\\&
\Gamma_{s} = \mbox{Tr}\left[ P_{(s)} \Gamma_{(s)} \right]\,,
\end{align}
where $\Gamma_{(v)}^{\mu}, \Gamma_{(p)}, \Gamma_{(a)}^{\mu}, \Gamma_{(s)}$    denote QCD amplitudes with tensor structures for the vector, pseudo-scalar, axial-vector, scalar currents, respectively.

Since contributions from soft, potential and ultrasoft loop momenta in full QCD and NRQCD are identical,  thus they drop out from both sides of the matching formula in Eq.~\eqref{matchformularenormalization}.
Then we set  $\widetilde{\Gamma}_{J}=1$~\cite{Marquard:2014pea,Feng:2022vvk} in Eq.~\eqref{matchformularenormalization} and   ${\Gamma}_{J}$  turns into  the hard  part of the full QCD, which entirely   determines   the matching coefficients  to all orders.
To match with NRQCD and obtain the matching coefficients, one need extract the contribution from the hard region in full QCD amplitudes for the heavy flavor-changing currents.
For the vector current, the pseudo-scalar current and the zeroth component of the axial-vector current case, we can simply set  the   small momentum  $k=0$ in the QCD amplitudes.
However, for  the space-like component of the axial-vector current and the scalar current  case, since NRQCD currents include contributions at $\mathcal{O}(k/m_r)$ from Eq.~\eqref{expandcurrents} and Eq.~\eqref{NRQCDcurrents}, accordingly,
one need first introduce the small  momentum $k$ to momenta in the QCD amplitudes  and then series expand propagator denominators with respect to $k$ up to $\mathcal{O}{(k)}$ in  the hard region of loop momenta~\cite{Kniehl:2006qw}, which will cancel $k^2$ in the denominators of the   projectors in Eq.~\eqref{projectorai} and Eq.~\eqref{projectors} so that we can obtain the finite final results   as $k^2 \rightarrow 0$.

Due to the expansion in powers of $k$, the number and powers of propagators in Feynman integrals constituting the amplitudes for the space-like component of the axial-vector current and the scalar current will remarkably increase compared with  the vector current case, the pseudo-scalar current and the zeroth component of the axial-vector current case. In our practice, the total number  of propagators in a three-loop Feynman integral family is 12 for the former two currents and 9 for the latter three currents. In our calculation, the most difficult and   time-consuming is
the reduction from three-loop Feynman integrals    with rank 5, dot 4, and 12  propagators   to master integrals. By trial and  error, we find it is more  appropriate for  Fire6~\cite{Smirnov:2019qkx} to deal with this problem than {\texttt{Kira}}~\cite{Klappert:2020nbg} or {\texttt{FiniteFlow}}~\cite{Peraro:2019svx}.
For  every heavy flavor-changing current, after  using  {\texttt{Kira}}+{\texttt{FIRE}}+{\texttt{Mathematica\,code}} to achieve the minimal set of master integral families  from  all of master integral families    based on symmetry among different integral families, the number of three-loop master integral families is reduced from 830s  to 26, meanwhile the number of three-loop master integrals is reduced from 13000s  to 300.

\section{NRQCD current renormalization constants~\label{ZjNRQCD}}

After implementing the quark field and  mass on-shell renormalization, and  the QCD coupling constant $\overline{\rm MS}$ renormalization, the QCD vertex function gets rid of ultra-violet (UV) poles, yet still contains  uncancelled infra-red (IR) poles starting from order $\alpha_s^2$.  The remaining IR poles in QCD should be exactly  cancelled by the UV divergences of ${\widetilde Z}_{J}$ in NRQCD,  rendering the matching coefficient finite.  With the aid of the obtained high-precision numerical results, combined with the structures and features of NRQCD  current renormalization constants investigated in other known literature~\cite{Feng:2022vvk,Feng:2022ruy,Sang:2022tnh,Egner:2022jot},    we have successfully reconstructed the exact analytical expressions of the NRQCD  renormalization constants for the heavy flavor-changing   currents
 through the PSLQ algorithm and 
numerical fitting recipes~\cite{Duhr:2019tlz}. Here we directly present the final results  as following
\begin{align}
\widetilde{Z}_{J}\left(x, {\mu^2_f\over m_b m_c } \right)&=1+\left(\frac{\alpha_{s}^{\left(n_{l}\right)}\left(\mu_f\right)}{\pi}\right)^{2}\widetilde{Z}_{J}^{(2)}\left(x\right)
 +\left(\frac{\alpha_{s}^{\left(n_{l}\right)}\left(\mu_f\right)}{\pi}\right)^{3}\widetilde{Z}_{J}^{(3)}\left(x, {\mu^2_f\over m_b m_c } \right)+\mathcal{O}(\alpha_s^4).
\end{align}
\begin{align}
\widetilde{Z}_{v}^{(2)}\left(x \right)&=\pi^{2}C_{F}\frac{1}{\epsilon}\left(\frac{3x^2+2x+3}{24\left(1+x\right)^2}C_{F}+\frac{1}{8}C_{A}\right),
\nonumber\\
\widetilde{Z}_{v}^{(3)}\left(x, {\mu^2_f\over m_b m_c } \right)&=
\pi^{2}C_{F}\bigg\{
C_F^2 \bigg[\frac{3 x^2-x+3}{36\epsilon ^2 (x+1)^2 }
+\frac{1}{\epsilon}\left(
\frac{19 x^2+5 x+19}{36 (x+1)^2}-\frac{2 }{3}\ln 2
\right.\nonumber\\&\left.
+\frac{x^3-4 x^2-2 x-3}{12 (x+1)^3}\ln x
+\frac{1}{6}
\ln (x+1)
	+\frac{3 x^2-x+3 }{12 (x+1)^2}\ln \frac{\mu _f^2}{m_b
		m_c}
\right)\bigg]
\nonumber\\ &
+
C_F C_A \bigg[
\frac{x}{216\epsilon ^2 (x+1)^2 }
+
\frac{1}{ \epsilon }\left(
\frac{39 x^2+148	x+39}{162 (x+1)^2}
\right.\nonumber\\&\left.
-\frac{x+11 }{48 (x+1)}\ln x
+\frac{1}{4} \ln (x+1)
+\frac{11 x^2+8 x+11}{48 (x+1)^2} \ln \frac{\mu _f^2}{m_b
	m_c}
\right)\bigg]
\nonumber\\ &
+C_A^2  \bigg[
\frac{-1}{16 \epsilon ^2}+\frac{1}{ \epsilon
}\left(
\frac{2}{27}+\frac{1}{6}\ln2-\frac{1}{24}\ln x+\frac{1}{12}\ln (x+1) +
\frac{1}{24}\ln \frac{\mu _f^2}{m_b	m_c}
\right)\bigg]
\nonumber\\ &
+ C_F T_F n_l\bigg[
\frac{3 x^2+2
	x+3}{108\epsilon ^2 (x+1)^2 }-\frac{21 x^2+58 x+21}{324 \epsilon (x+1)^2 }
\bigg]
+C_A  T_F n_l \bigg[
\frac{1}{36 \epsilon ^2}-\frac{37}{432 \epsilon }
\bigg]
\nonumber\\ &
+C_F  T_F n_b
\frac{ x^2 }{15\epsilon (x+1)^2  }
+C_F  T_F n_c
\frac{ 1 }{15\epsilon (x+1)^2  }
\bigg\}.
\end{align}
\begin{align}
\widetilde{Z}_{p}^{(2)}\left(x \right)&= \pi^{2}C_{F}\frac{1}{\epsilon}\left(\frac{x^2+6 x+1}{8 (x+1)^2}C_{F}+\frac{1}{8}C_{A}\right),
\nonumber\\
\widetilde{Z}_{p}^{(3)}\left(x, {\mu^2_f\over m_b m_c } \right)&=
\pi^{2}C_{F}\bigg\{
C_F^2 \bigg[
\frac{3 x^2-x+3}{36\epsilon ^2 (x+1)^2 }
+\frac{1}{\epsilon}\left(\frac{19 x^2+29 x+19}{36 (x+1)^2}-\frac{2 }{3}\ln 2
\right.\nonumber\\&\left.
+\frac{x^4-5 x^3-22 x^2-x+3 }{12 (x-1) (x+1)^3}\ln x+\frac{1}{6} \ln (x+1)
+\frac{3	x^2-x+3}{12 (x+1)^2}\ln \frac{\mu _f^2}{m_b	m_c}
\right)\bigg]
\nonumber\\ &
+
C_F C_A \bigg[
\frac{-5 x}{24\epsilon ^2 (x+1)^2 }
+
\frac{1}{ \epsilon }\left(
\frac{26 x^2+93	x+26}{108 (x+1)^2}
\right.\nonumber\\&\left.
+\frac{-x^2+2 x+11 }{48 (x-1) (x+1)}\ln x+\frac{1}{4} \ln (x+1)+
 \frac{11 x^2+36 x+11 }{48 (x+1)^2}\ln \frac{\mu _f^2}{m_b	m_c}
\right)\bigg]
\nonumber\\ &
+C_A^2  \bigg[
\frac{-1}{16 \epsilon ^2}+\frac{1}{ \epsilon
}\left(
\frac{2}{27}+\frac{1}{6}\ln 2-\frac{1}{24}\ln x+\frac{1}{12} \ln (x+1)+\frac{1}{24} \ln \frac{\mu _f^2}{m_b	m_c}
\right)\bigg]
\nonumber\\ &
+ C_F T_F n_l\bigg[
\frac{x^2+6 x+1}{36 \epsilon ^2 (x+1)^2}-\frac{7 x^2+30 x+7}{108 \epsilon 	(x+1)^2}
\bigg]
+C_A  T_F n_l \bigg[
\frac{1}{36 \epsilon ^2}-\frac{37}{432 \epsilon }
\bigg]
\nonumber\\ &
+C_F  T_F n_b
\frac{ x^2 }{15\epsilon (x+1)^2  }
+C_F  T_F n_c
\frac{ 1 }{15\epsilon (x+1)^2  }
\bigg\}.
\end{align}
\begin{align}
\widetilde{Z}_{a,i}^{(2)}\left(x \right)&= \pi^{2}C_{F}\frac{1}{\epsilon}\left(\frac{3x^2+4x+3}{24\left(1+x\right)^2}C_{F}+\frac{1}{24}C_{A}\right),
\nonumber\\
\widetilde{Z}_{a,i}^{(3)}\left(x, {\mu^2_f\over m_b m_c } \right)&=
\pi^{2}C_{F}\bigg\{
\frac{C_F^2}{\epsilon}\bigg[\frac{171 x^2-296 x+171}{216
	(x+1)^2}-\frac{\ln 2}{3}
\nonumber\\ &
+\frac{-57 x^4+89 x^3+274 x^2+89 x-57}{216 (x-1) (x+1)^3} \ln x\bigg]
\nonumber\\ &
+
C_F C_A \bigg[	-\frac{22 x^2+31	x+22}{432\epsilon ^2 (x+1)^2 }
+
\frac{1}{ \epsilon }\left(
\frac{379 x^2+675 x+379}{1296 (x+1)^2}-\frac{\ln 2}{18}
\right.\nonumber\\&\left.
-\frac{5 x+11}{144 (x+1)} \ln x+\frac{1}{9} \ln (x+1)
+\frac{11 x^2+13 x+11}{144	(x+1)^2} \ln \frac{\mu _f^2}{m_b
	m_c}
\right)\bigg]
\nonumber\\ &
+C_A^2  \bigg[\frac{-1}{48 \epsilon ^2}+\frac{1}{648 \epsilon
}\left(34+72 \ln2-9 \ln x+18 \ln (x+1)+9 \ln \frac{\mu _f^2}{m_b m_c} \right)\bigg]
\nonumber\\ &
+ C_F T_F n_l\bigg[
\frac{3 x^2+4
	x+3}{108\epsilon ^2 (x+1)^2 }-\frac{21 x^2+41 x+21}{324 \epsilon (x+1)^2 }
\bigg]
\nonumber\\ &
+C_A  T_F n_l \bigg[\frac{1}{108 \epsilon ^2}-\frac{53}{1296 \epsilon }\bigg] \bigg\}.
\end{align}
\begin{align}
\widetilde{Z}_s^{(2)}\left(x \right)&= \pi^{2}C_{F}\frac{1}{\epsilon}\left(\frac{3x^2+10x+3}{24\left(1+x\right)^2}C_{F}+\frac{1}{24}C_{A}\right),
\nonumber\\
\widetilde{Z}_s^{(3)}\left(x, {\mu^2_f\over m_b m_c } \right)&=
\pi^{2}C_{F}\bigg\{\frac{C_F^2}{\epsilon}\bigg[\frac{57x^2+146x+57}{216(x+1)^2}-\frac{\ln2}{3}\bigg]
\nonumber\\ &
+
C_F C_A \bigg[-\frac{11 x^2+41 x+11}{216\epsilon ^2	(x+1)^2 }
+
\frac{1}{1296 \epsilon }\left(\frac{379 x^2+1086 x+379}{(x+1)^2}-72 \ln2
\right.\nonumber\\&\left.
-\frac{9 (5 x+11) }{x+1}\ln x+144 \ln (x+1)+\frac{9 \left(11 x^2+28 x+11\right) }{(x+1)^2}\ln \frac{\mu _f^2}{m_b
	m_c}
\right)\bigg]
\nonumber\\ &
+C_A^2  \bigg[\frac{-1}{48 \epsilon ^2}+\frac{1}{648 \epsilon
}\left(34+72 \ln2-9 \ln x+18 \ln (x+1)+9 \ln \frac{\mu _f^2}{m_b m_c} \right)\bigg]
\nonumber\\ &
+ C_F T_F n_l\bigg[\frac{3 x^2+10
	x+3}{108\epsilon ^2 (x+1)^2 }-\frac{21 x^2+74 x+21}{324 \epsilon (x+1)^2 }\bigg]
\nonumber\\ &
+C_A  T_F n_l \bigg[\frac{1}{108 \epsilon ^2}-\frac{53}{1296 \epsilon }\bigg] \bigg\},
\end{align}
where $C_A=3$, $C_F=4/3$ and $T_F=1/2$ are QCD constants and we have defined a dimensionless parameter $x$ representing the ratio of two heavy quark masses
\begin{align}
x= {m_c\over m_b}.
\end{align}

Then the corresponding anomalous dimension $\tilde{\gamma}_{J}$ for the heavy flavor-changing   NRQCD  currents  is related to $\widetilde{Z}_{J}$ by~\cite{Groote:1996xb,Kiselev:1998wb,Henn:2016tyf,Fael:2022miw,Grozin:2015kna,Ozcelik:2021zqt}
\begin{align}
\tilde{\gamma}_{J}\left(x, {\mu^2_f\over m_b m_c } \right) &\equiv
{d \ln \widetilde{Z}_{J} \over d \ln \mu_f }  
\equiv
\frac{-2\, \partial{\widetilde{Z}_{J}^{[1]}}}{\partial\ln\alpha_s^{(n_l)}(\mu_f)}
\nonumber\\ &
 =\left(\frac{\alpha_s^{(n_l)}
	\left(\mu_f\right)}{\pi}\right)^2 \tilde{\gamma}_{J}^{(2)}(x )
+\left(\frac{\alpha_s^{\left(n_l\right)
	}\left(\mu_f\right)}{\pi}\right)^3\tilde{\gamma}_{J}^{(3)}\left(x, {\mu^2_f\over m_b m_c }\right)+\mathcal{O}(\alpha^4_s),
\end{align}
where $\widetilde{Z}_{J}^{[1]}$ denotes the coefficient of the $\frac{1}{\epsilon}$ pole in $\widetilde{Z}_{J}$, and within the NRQCD frame we use $\mu_f$ to denote the  NRQCD factorization scale, 
 so that both  $\widetilde{Z}_{J}$ and  $\tilde{\gamma}_{J}$ explicitly depend on  $\mu_f$ but not the QCD renormalization scale $\mu$~\cite{Marquard:2014pea,Egner:2022jot,Feng:2022vvk,Feng:2022ruy,Sang:2022tnh}. Explicitly, $\tilde{\gamma}_{J}^{(2)}$ and $\tilde{\gamma}_{J}^{(3)}$ read
\begin{align}
\tilde{\gamma}_{v}^{(2)}\left(x \right)&=-\pi^2C_F \bigg[C_F \frac{3 x^2+2	x+3}{6(x+1)^2}+\frac{C_A}{2} \bigg],
\nonumber\\
\tilde{\gamma}_{v}^{(3)}\left(x, {\mu^2_f\over m_b m_c } \right)&=
\pi^{2}C_{F} \bigg\{
C_F^2 \bigg[
-\frac{19 x^2+5 x+19}{6 (x+1)^2}+4 \ln 2
\nonumber\\&
+\frac{-x^3+4 x^2+2 x+3}{2 (x+1)^3}\ln x-\ln (x+1)
+\frac{-3 x^2+x-3 }{2 (x+1)^2}\ln\frac{\mu_f^2}{m_b m_c}
\bigg]
\nonumber\\ &
+C_F C_A \bigg[
-\frac{39 x^2+148 x+39}{27	(x+1)^2}+\frac{x+11 }{8( x+1)}\ln x
\nonumber\\ &
-\frac{3}{2} \ln (x+1)
-\frac{11 x^2+8 x+11}{8 (x+1)^2}\ln\frac{\mu_f^2}{m_b m_c}
\bigg]
\nonumber\\ &	
+C_A^2 \bigg[
-\frac{4}{9}-\ln 2+\frac{1}{4}\ln x-\frac{1}{2} \ln (x+1)
-\frac{1}{4} \ln \frac{\mu_f^2}{m_b m_c}
 \bigg]
 \nonumber\\ &
+C_F T_F n_l \frac{21 x^2+58 x+21}{54 (x+1)^2}
+\frac{37}{72} C_A T_F n_l
\nonumber\\ &	
-\frac{2 x^2 }{5 (x+1)^2} C_F T_F n_b
-\frac{2}{5 (x+1)^2} C_F T_F n_c
\bigg\}.
\end{align}
\begin{align}
\tilde{\gamma}_{p}^{(2)}\left(x \right)&=
-\pi^2C_F \bigg[\frac{x^2+6 x+1 }{2 (x+1)^2}C_F+\frac{C_A }{2} \bigg],
\nonumber\\
\tilde{\gamma}_{p}^{(3)}\left(x, {\mu^2_f\over m_b m_c } \right)&=
\pi^{2}C_{F} \bigg\{
C_F^2 \bigg[
-\frac{19 x^2+29 x+19}{6 (x+1)^2}+4 \ln 2
\nonumber\\ &
+\frac{-x^4+5 x^3+22 x^2+x-3 }{2 (x-1)	(x+1)^3}\ln x-\ln (x+1)+\frac{-3 x^2+x-3}{2 (x+1)^2}\ln\frac{\mu_f^2}{m_b m_c}
\bigg]
\nonumber\\ &
+C_F C_A \bigg[
-\frac{26 x^2+93 x+26}{18	(x+1)^2}-\frac{-x^2+2 x+11 }{8(x-1)(x+1)}\ln x
\nonumber\\ &
-\frac{3}{2} \ln (x+1)
-\frac{11 x^2+36 x+11}{8 (x+1)^2} \ln\frac{\mu_f^2}{m_b m_c}
\bigg]
\nonumber\\ &	
+C_A^2 \bigg[
-\frac{4}{9}-\ln 2+\frac{1}{4}\ln x-\frac{1}{2} \ln (x+1)
-\frac{1}{4} \ln \frac{\mu_f^2}{m_b m_c}
\bigg]
\nonumber\\ &
+C_F T_F n_l \frac{7 x^2+30	x+7}{18 (x+1)^2}
+\frac{37}{72} C_A T_F n_l
\nonumber\\ &	
-\frac{2 x^2 }{5 (x+1)^2} C_F T_F n_b
-\frac{2}{5 (x+1)^2} C_F T_F n_c
\bigg\}.
\end{align}
\begin{align}
\tilde{\gamma}_{a,i}^{(2)}\left(x \right)&=
-\pi^2C_F \bigg[C_F \frac{3 x^2+4	x+3}{6(x+1)^2}+\frac{C_A}{6} \bigg],
\nonumber\\
\tilde{\gamma}_{a,i}^{(3)}\left(x, {\mu^2_f\over m_b m_c } \right)&=
\pi^{2}C_{F} \bigg\{
C_F^2 \bigg[\frac{-171 x^2+296 x-171}{36 (x+1)^2}+2 \ln 2
\nonumber\\ &
+\frac{57 x^4-89 x^3-274 x^2-89 x+57 }{36 (x-1) (x+1)^3}\ln x\bigg]
\nonumber\\ &
+C_F C_A \bigg[-\frac{379 x^2+675 x+379}{216 (x+1)^2}+\frac{\ln 2}{3}
\nonumber\\ &
+\frac{5 x+11}{24( x+1)}\ln x
-\frac{2}{3} \ln (x+1)
	-\frac{11 x^2+13 x+11}{24 (x+1)^2}\ln\frac{\mu
		_f^2}{m_b m_c}\bigg]
\nonumber\\ &	
+C_A^2 \bigg[
-\frac{17}{54}-\frac{2 }{3}\ln 2+\frac{1}{12}\ln x-\frac{1}{6} \log (x+1)-\frac{1}{12} \ln \frac{\mu_f^2}{m_b m_c}\bigg]
\nonumber\\ &
+C_F T_F n_l \frac{21 x^2+41 x+21}{54 (x+1)^2}
+\frac{53}{216} C_A T_F n_l \bigg\}.
\end{align}
\begin{align}
\tilde{\gamma}_s^{(2)}\left(x \right)&=
-\pi^2C_F \bigg[C_F \frac{3 x^2+10	x+3}{6(x+1)^2}+\frac{C_A}{6} \bigg],
\nonumber\\
\tilde{\gamma}_s^{(3)}\left(x, {\mu^2_f\over m_b m_c } \right)&=
\pi^{2}C_{F} \bigg\{
C_F^2 \bigg[-\frac{57 x^2+146 x+57}{36 (x+1)^2}+2 \ln 2\bigg]
\nonumber\\ &
+C_F C_A \bigg[-\frac{379 x^2+1086 x+379}{216 (x+1)^2}
+\frac{\ln 2}{3}
\nonumber\\ &
+\frac{5 x+11 }{24 (x+1)}\ln x-\frac{2}{3} \ln (x+1)-\frac{11 x^2+28 x+11}{24 (x+1)^2}\ln\frac{\mu
	_f^2}{m_b m_c}\bigg]
\nonumber\\ &	
+C_A^2 \bigg[-\frac{17}{54}-\frac{2 }{3}\ln 2+\frac{1}{12}\ln x-\frac{1}{6} \ln (x+1)-\frac{1}{12} \ln \frac{\mu_f^2}{m_b m_c}\bigg]
\nonumber\\ &
+C_F T_F n_l \frac{21 x^2+74 x+21}{54 (x+1)^2}
+\frac{53}{216} C_A T_F n_l \bigg\}.
\end{align}
Note that the above analytical expressions of $\widetilde{Z}_{a,i}$ and $\tilde{\gamma}_{a,i}$ for the space-like  component  of the heavy flavor-changing  axial-vector   current  are new.
The obtained $\widetilde{Z}_{J}$ and $\tilde{\gamma}_{J}$ have been checked with several different values of $m_b$ and $m_c$.
To verify the correctness of our results, on the one hand, one can check that the above $\widetilde{Z}_{J}$ (also $\tilde{\gamma}_{J}$) is  symmetric/invariant under the exchange $m_b\leftrightarrow m_c$ meanwhile $n_b\leftrightarrow n_c$, on the other hand, in the equal quark masses case $x=1$, our $\widetilde{Z}_{J}$ and $\tilde{\gamma}_{J}$ are in full agreement with the known results in Refs.~\cite{Kniehl:2006qw,Piclum:2007an,Egner:2022jot}.

In our calculation, we include the contributions from the loops of charm quark and bottom quark in full QCD, which however are decoupled in the NRQCD.
To match QCD with NRQCD, one need twice apply the decoupling relation~\cite{Chetyrkin:2005ia,Bernreuther:1981sg,Grozin:2011nk,Barnreuther:2013qvf,Grozin:2007fh,Gerlach:2019kfo,Ozcelik:2021zqt} of  $\alpha_s$ :
\begin{align}\label{Decoupling}
\frac{\alpha_s^{(n_l+1)}(\mu)}{\alpha_s^{(n_l)}(\mu)} =& 1-\frac{\alpha_s^{(n_l)}(\mu)}{\pi}\frac{
	(1-I_0 \epsilon )}{3  \epsilon }T_F
+
\left( \frac{\alpha_s^{(n_l)}(\mu)}{\pi} \right)^2
 T_F \Bigg\{C_A \left(\frac{I_0^2 \epsilon  \left(4 \epsilon ^3+4 \epsilon ^2-11 \epsilon -10\right)}{8
		(\epsilon -2) (2 \epsilon +1) (2 \epsilon +3)}
\right.	\nonumber\\&\left.	
	-\frac{5}{24 \epsilon }\right)	
	+C_F \left(-\frac{I_0^2 \epsilon  \left(4 \epsilon ^3-7 \epsilon -1\right)}{4
		(\epsilon -2) (2 \epsilon -1) (2 \epsilon +1)}-\frac{1}{8 \epsilon }\right)+T_F \frac{(1-I_0 \epsilon )^2}{9 \epsilon ^2}\Bigg\}
+ \mathcal{O}{(\alpha_s^3)},
\end{align}
where $I_0= e^{\gamma_E  \epsilon } (\epsilon -1) \Gamma (\epsilon -1) {\left({\mu
	^2}/{m_Q^2}\right)}^{\epsilon }$ and $m_Q$ is the on-shell mass of the decoupled heavy
quark.

Besides, we  can evolve the strong coupling from the scale $\mu_f$ to the scale $\mu$ with  renormalization group running equation~\cite{Abreu:2022cco} in $D=4-2 \epsilon$ dimensions  as following
\begin{align}\label{asrun1}
\alpha_s^{(n_l)}\left(\mu_f\right)=
\alpha_s^{(n_l)}\left(\mu\right)\left(\frac{\mu}{\mu_f}\right)^{2\epsilon}
\bigg[1+\frac{\alpha_s^{(n_l)}\left(\mu\right)}{\pi}\frac{\beta_0^{(n_l)}}{4\epsilon}\left(\left(\frac{\mu}{\mu_f}\right)^{2\epsilon}-1\right)+ \mathcal{O}{(\alpha_s^2)}\bigg] .
\end{align}
To calculate the values of the strong coupling  constant  $\alpha_s$,  we  also use the renormalization group running equation~\cite{Chetyrkin:2000yt} in $D=4$ dimensions as
\begin{align}\label{asrun2}
\alpha_s^{(n_l)}\left(\mu\right)=
\frac{4\pi}{\beta_0^{(n_l)}L_{\Lambda}}\Bigg[1-\frac{b_1 \ln L_{\Lambda}}{\beta_0^{(n_l)}L_{\Lambda}}+\frac{b_1^2(\ln^2 L_{\Lambda}-\ln L_{\Lambda}-1)+b_2}{\left(\beta_0^{(n_l)}L_{\Lambda}\right)^2}+\mathcal{O}{\left(\left(\frac{1}{L_{\Lambda}}\right)^3\right)}\Bigg],
\end{align}
where $L_{\Lambda}=\ln\left(\mu^2/{\Lambda_{QCD}^{(n_l)}}^2\right)$,  $b_i=\beta_i^{(n_l)}/{\beta_0^{(n_l)}}$. And $\beta_0$, $\beta_1$,  $\beta_2$ are the one-loop, two-loop and three-loop coefficients of
the QCD $\beta$ function~\cite{vanRitbergen:1997va}, respectively, which read
\begin{align}
&\beta_0^{(n_l)}=\frac{11}{3}C_A-\frac{4}{3} T_F n_l,
\nonumber\\&
\beta_1^{(n_l)}=\frac{34}{3}C_A^2-\frac{20}{3} C_A T_F n_l-4 C_F T_F n_l,
\nonumber\\&
\beta_2^{(n_l)}=\frac{2857}{54}C_A^3-\left(\frac{1415}{27} C_A^2+\frac{205}{9} C_A C_F-2C_F^2\right) T_F n_l+\left(\frac{158}{27} C_A+\frac{44}{9}  C_F\right) T_F^2 n_l^2.
\end{align}
In our numerical evaluation,  $n_b=n_c=1$, $n_l=3$ are fixed through the decoupling region from $\mu=0.4\,\mathrm{GeV}$ to $\mu=7\,\mathrm{GeV}$,
and the typical QCD scale $\Lambda_{QCD}^{(n_l=3)}=0.3344\mathrm{GeV}$  is  determined using three-loop formula with the aid of the package {\texttt{RunDec}}~\cite{Chetyrkin:2000yt,Schmidt:2012az,Deur:2016tte,Herren:2017osy}  by inputting the initial value  $\alpha_s^{(n_f=5)}\left(m_Z=91.1876\mathrm{GeV}\right)=0.1179$.

\section{Matching coefficients~\label{Matching coefficients}}
Following Refs.~\cite{Feng:2022vvk,Feng:2022ruy,Sang:2022tnh}, the dimensionless matching coefficient $\mathcal{C}_J$ for the heavy flavor-changing  currents  can be decomposed as:
\begin{align}
\label{Cjformula}
& \mathcal{C}_J(\mu_f,\mu,m_b,m_c) =1+\frac{\alpha_s^{\left(n_l\right)}\left(\mu\right)}{\pi} \mathcal{C}_J^{(1)}(x)
+\left(\frac{\alpha_s^{\left(n_l\right)}\left(\mu\right)}{\pi}\right)^2
\left(\mathcal{C}_J^{(1)}(x)\frac{\beta_0^{(n_l)}}{4}\text{ln}\frac{\mu^2}{m_b m_c}
\right.\nonumber \\&\left.
+\frac{\tilde{\gamma}_J^{(2)}(x)}{2}\ln \frac{\mu_f^2}{m_b m_c}+\mathcal{C}_J^{(2)}(x)\right)
 +\left(\frac{\alpha_s^{\left(n_l\right)}\left(\mu\right)}{\pi}\right)^3\Bigg\{ \left(\frac{\mathcal{C}_J^{(1)}(x)}{16}\beta_1^{(n_l)}+\frac{\mathcal{C}_J^{(2)}(x)}{2}\beta_0^{(n_l)}\right)\text{ln}\frac{\mu^2}{m_b m_c}
\nonumber \\&
+\frac{\mathcal{C}_J^{(1)}(x)}{16}{\beta_0^{(n_l)}}^2\ln^2
\frac{\mu^2}{m_b m_c}
+\frac{1}{8}\left(\frac{d\tilde{\gamma}_J^{(3)}\left(x,\frac{\mu_f^2}{m_b m_c}\right)}{d \text{ln}\mu_f}-\beta_0^{(n_l)}\tilde{\gamma}_J^{(2)}(x)\right)\ln^2\frac{\mu_f^2}{m_b m_c}  
\nonumber \\&
+\frac{1}{2}\left(\mathcal{C}_J^{(1)}(x) \tilde{\gamma}_J^{(2)}(x)+\tilde{\gamma}_J^{(3)}\left(x,1\right)\right)\ln\frac{\mu_f^2}{m_b m_c}
+\frac{\beta_{0}^{(n_l)}}{4}\tilde{\gamma}_J^{(2)}(x)\ln\frac{\mu_f^2}{m_b m_c}\,~\text{ln}\frac{\mu^2}{m_b m_c}
+ \mathcal{C}_J^{(3)}(x) \Bigg\}
\nonumber \\&
+\mathcal{O}\left(\alpha_s^4\right),
\end{align}
where $n_l=3$ denotes the massless flavours. $\mathcal{C}_J^{(n)}(x)(n=1,2,3)$, as a function of $x=m_c/m_b$, independent of $\mu$ and $\mu_f$, corresponds to the nontrivial part at $\mathcal{O}\left(\alpha_s^n\right)$ of $\mathcal{C}_J$.
It's well known that $\mathcal{C}_J$ and $\mathcal{C}_J^{(n)}(x)$ satisfy the following symmetry properties~\cite{Braaten:1995ej,Hwang:1999fc,Lee:2010ts,Onishchenko:2003ui,Chen:2015csa,Feng:2022ruy,Sang:2022tnh}:
\begin{align}\label{cjsym}
&\mathcal{C}_J(\mu_f,\mu,m_b,m_c)=\mathcal{C}_J(\mu_f,\mu,m_c,m_b)|_{\,  n_b\leftrightarrow n_c},
\nonumber\\&
\mathcal{C}_J^{(n)}(x)=\mathcal{C}_J^{(n)}\left(\frac{1}{x}\right)|_{\,  n_b\leftrightarrow n_c}.
\end{align}

The nontrivial one-loop piece  $\mathcal{C}_J^{(1)}(x)$ can be analytically achieved as:
\begin{align}
&\mathcal{C}_v^{(1)}(x)=\frac{3}{4}C_F\left(\frac{x-1}{x+1}\,\ln x-\frac{8}{3}\right),
\\&
\mathcal{C}_p^{(1)}(x)=\frac{3}{4}C_F\left(\frac{x-1}{x+1}\,\ln x-2\right),
\\&
\mathcal{C}_{a,i}^{(1)}(x)=\frac{3}{4} C_F \left(\frac{x-1}{x+1}\,\ln x-\frac{4}{3}\right),
\\&
\mathcal{C}_s^{(1)}(x)=\frac{3}{4} C_F \left(\frac{x-1}{x+1}\,\ln x-\frac{2}{3}\right).
\end{align}

And the nontrivial two-loop and three-loop pieces in Eq.~\eqref{Cjformula} are  $\mathcal{C}_J^{(2)}(x)$ and $\mathcal{C}_J^{(3)}(x)$, respectively, which following the convention of Refs. \cite{Marquard:2014pea,Beneke:2014qea,Egner:2022jot,Feng:2022vvk,Feng:2022ruy,Sang:2022tnh}, can be further decomposed  in terms of different color/flavor structures:
\begin{align}
\mathcal{C}_J^{(2)}(x) =& C_F \, C_F \,\mathcal{C}_J^{FF}(x)+C_F \, C_A\, \mathcal{C}_J^{FA}(x)
\nonumber \\&
+C_F \, T_F\, n_b \,\mathcal{C}_J^{FB}(x) +C_F \, T_F\, n_c\, \mathcal{C}_J^{FC}(x) +C_F \, T_F\, n_l\, \mathcal{C}_J^{FL}(x),
\\
\mathcal{C}_J^{(3)}(x) =&   C^3_F \, \mathcal{C}_J^{FFF}(x)+C_F^2 \,C_A \, \mathcal{C}_J^{FFA}(x)
+C_F\, C_A^2 \, \mathcal{C}_J^{FAA}(x)
\nonumber \\&
 +C_F^2\, T_F\, n_l\, \mathcal{C}_J^{FFL}(x) +C_F\,C_A\,T_F\, n_l\,\mathcal{C}_J^{FAL}(x)
+C_F\,T_F^2 \, n_c \, n_l\,  \mathcal{C}_J^{FCL}(x)   
 \nonumber\\&
+C_F\,T_F^2 \, n_b \, n_l\,  \mathcal{C}_J^{FBL}(x) +C_F\,T_F^2\, n_l^2\, \mathcal{C}_J^{FLL}(x)
    +C_F\, T_F^2 \, n_b \, n_c \, \mathcal{C}_J^{FBC}(x)
 \nonumber\\&
    +
C_F^2\,T_F \, n_c\,  \mathcal{C}_J^{FFC}(x)
    +C_F\, C_A\,T_F \, n_c\, \mathcal{C}_J^{FAC}(x)+
 C_F\, T_F^2 \, n_c^2\,  \mathcal{C}_J^{FCC}(x)
\nonumber\\
&  + C_F^2\,  T_F\, n_b\, \mathcal{C}_J^{FFB}(x)+C_F\, C_A\,T_F\, n_b\, \mathcal{C}_J^{FAB}(x)+C_F\,T_F^2\, n_b^2\,
 \mathcal{C}_J^{FBB}(x).
\end{align}

Due to limited computing resources,
we choose to calculate the matching coefficient $\mathcal{C}_J$ at three rational numerical points: the physical point  $\{m_b=\frac{475}{100}\,\mathrm{GeV}, m_c=\frac{150}{100}\,\mathrm{GeV}\}$ $\left(i.e.,x=x_0=\frac{150}{475}\right)$,   the check  point $x=1/x_0$ and the equal mass point $x=1$, respectively.
The results $\mathcal{C}_J$ obtained at the physical point and the check  point verify the symmetric features of $\mathcal{C}_J$ and $\mathcal{C}_J^{(n)}(x)$ in Eq.~\eqref{cjsym}. Our results $\mathcal{C}_J$ obtained at the equal mass point $x=1$ are consistent with the known results $\mathcal{C}_J$ for all four currents in the equal quark masses case in the literature~\cite{Egner:2022jot,Marquard:2014pea,Kniehl:2006qw,Piclum:2007an}. Furthermore, our calculation  have verified   the three-loop matching coefficients  for the  heavy flavor-changing    pseudoscalar current  and   the zeroth component of  the heavy flavor-changing   axial-vector current are identical, i.e., $\mathcal{C}_p\equiv\mathcal{C}_{a,0}$.
To confirm our calculation,  we have also calculated   $\mathcal{C}_p$ and  $\mathcal{C}_v$ at the reference point $\{m_b=\frac{498}{100}\,\mathrm{GeV}, m_c=\frac{204}{100}\,\mathrm{GeV}\}$,  where  our results   agree with the known results in Refs.~\cite{Feng:2022ruy,Sang:2022tnh}.

In the following, we will present the highly accurate numerical results of $\mathcal{C}_J^{(2)}(x)$ and $\mathcal{C}_J^{(3)}(x)$ at the physical heavy quark mass ratio  $x=x_0=\frac{150}{475}$  with about 30-digit precision.
The various color-structure components of ${\cal C}_{v}^{(2)}(x_0)$ and  ${\cal C}_{v}^{(3)}(x_0)$  read:
\begin{subequations}
	\begin{align}
	&\mathcal{C}_{v}^{FF}(x_0) = -13.71289080533129643353786882415,
	\\&
	\mathcal{C}_{v}^{FA}(x_0) = -6.585499135192203408065908804167,
	\\&
	\mathcal{C}_{v}^{FB}(x_0) = 0.0947676481125652606487968503976,
	\\&
	\mathcal{C}_{v}^{FC}(x_0) = 0.585796563729044305159251023619,
	\\&
	\mathcal{C}_{v}^{FL}(x_0) = 0.486237497534452686364818186481.
\\	&
 \mathcal{C}_{v}^{FFF}(x_0) =20.18969417129305999911571842286 ,
	\\&
	\mathcal{C}_{v}^{FFA}(x_0) = -203.4349264860295194232572876813,
	\\&
	\mathcal{C}_{v}^{FAA}(x_0) = -102.7968727737777422224763578788,
	\\&
	\mathcal{C}_{v}^{FFL}(x_0) = 50.93775016890326146248907065956,
	\\&
	\mathcal{C}_{v}^{FAL}(x_0) = 40.22574662383519955538190917802,
	\\&
	\mathcal{C}_{v}^{FCL}(x_0) = -0.776339576123527777867508257477,
	\\&
	\mathcal{C}_{v}^{FBL}(x_0) =-0.0556259617628169261333544284783 ,
	\\&
	\mathcal{C}_{v}^{FLL}(x_0) = -2.08814878247962216692347776970,
	\\&
	\mathcal{C}_{v}^{FBC}(x_0) = 0.0903048438843974616499880474411,
	\\&
	\mathcal{C}_{v}^{FFC}(x_0) = -1.6854789447153670526748653364,
	\\&
	\mathcal{C}_{v}^{FAC}(x_0) = 0.464663487323886298396191419941,
	\\&
	\mathcal{C}_{v}^{FCC}(x_0) = 0.166410566769625472334622650374,
	\\&
	\mathcal{C}_{v}^{FFB}(x_0) = -0.125493504901815435721244899040,
	\\&
	\mathcal{C}_{v}^{FAB}(x_0) = -0.207735042283005003179604843189,
	\\&
	\mathcal{C}_{v}^{FBB}(x_0) = 0.0155302263395316874159466507910.
	\end{align}
\end{subequations}
The various color-structure components of ${\cal C}_{p}^{(2)}(x_0)$ and  ${\cal C}_{p}^{(3)}(x_0)$  read:
\begin{subequations}
	\begin{align}
	&\mathcal{C}_{p}^{FF}(x_0) = -13.2664163032173887184051674623,
	\\&
	\mathcal{C}_{p}^{FA}(x_0) =-8.01693280168379244869854709253,
	\\&
	\mathcal{C}_{p}^{FB}(x_0) =0.128918313022470380156209778800 ,
	\\&
	\mathcal{C}_{p}^{FC}(x_0) = 0.709389226649688045706906749579,
	\\&
	\mathcal{C}_{p}^{FL}(x_0) =0.0583586106180016483840747484073 .
\\	&
  \mathcal{C}_{p}^{FFF}(x_0) = -19.1073059443195339789160242369,
	\\&
	\mathcal{C}_{p}^{FFA}(x_0) = -199.368217320524804838849517451,
	\\&
	\mathcal{C}_{p}^{FAA}(x_0) = -108.136868017309597645533701505,
	\\&
	\mathcal{C}_{p}^{FFL}(x_0) = 54.7202915496511304152624843564,
	\\&
	\mathcal{C}_{p}^{FAL}(x_0) = 40.1950559961352476805071968834,
	\\&
	\mathcal{C}_{p}^{FCL}(x_0) = -0.72340116280534841918416586699,
	\\&
	\mathcal{C}_{p}^{FBL}(x_0) = -0.019622466229502871214841024261,
	\\&
	\mathcal{C}_{p}^{FLL}(x_0) = -1.2331473244200630628135034057,
	\\&
	\mathcal{C}_{p}^{FBC}(x_0) = 0.157356536599972434859830922842,
	\\&
	\mathcal{C}_{p}^{FFC}(x_0) =4.93713785637736911159215602248 ,
	\\&
	\mathcal{C}_{p}^{FAC}(x_0) = -0.837019369825724302408030392435,
	\\&
	\mathcal{C}_{p}^{FCC}(x_0) = 0.256031525456388447241169171007,
	\\&
	\mathcal{C}_{p}^{FFB}(x_0) = 1.74108819472094062081877632041,
	\\&
	\mathcal{C}_{p}^{FAB}(x_0) = -0.675630407239784126013977083266,
	\\&
	\mathcal{C}_{p}^{FBB}(x_0) = 0.030203809353562550021046660214.
	\end{align}
\end{subequations}
The various color-structure components of ${\cal C}_{a,i}^{(2)}(x_0)$ and  ${\cal C}_{a,i}^{(3)}(x_0)$  read:
\begin{subequations}
\begin{align}
&\mathcal{C}_{a,i}^{FF}(x_0) = -7.55810100985517328528944527445,
\\&
\mathcal{C}_{a,i}^{FA}(x_0) = -4.00029433499832597188660997837,
\\&
\mathcal{C}_{a,i}^{FB}(x_0) = 0.01402013116548156662918961686909,
\\&
\mathcal{C}_{a,i}^{FC}(x_0) = 0.1451892135460458985527809307761,
\\&
\mathcal{C}_{a,i}^{FL}(x_0) = 0.06421749358957698454852728485104.
\\	&
 \mathcal{C}_{a,i}^{FFF}(x_0) = -8.42643917300377651280522219395,
	\\&
	\mathcal{C}_{a,i}^{FFA}(x_0) = -93.2038302986321307103761492345,
	\\&
	\mathcal{C}_{a,i}^{FAA}(x_0) = -63.7233070196452524188390331877,
	\\&
	\mathcal{C}_{a,i}^{FFL}(x_0) = 30.26464969513072132366156126441,
	\\&
	\mathcal{C}_{a,i}^{FAL}(x_0) = 21.16033670918183535060836071864,
	\\&
	\mathcal{C}_{a,i}^{FCL}(x_0) = -0.3153222090952701782698048799971,
	\\&
	\mathcal{C}_{a,i}^{FBL}(x_0) = -0.04600893146925268206019173772344,
	\\&
	\mathcal{C}_{a,i}^{FLL}(x_0) = -0.5156996834396487701112794535693,
	\\&
	\mathcal{C}_{a,i}^{FBC}(x_0) = -0.0118687507006037939590736816982,
	\\&
	\mathcal{C}_{a,i}^{FFC}(x_0) = 1.972145133214113574301216956655,
	\\&
	\mathcal{C}_{a,i}^{FAC}(x_0) = -0.2648389457862748902646421554436,
	\\&
	\mathcal{C}_{a,i}^{FCC}(x_0) = 0.005739492344304311786485592038925,
	\\&
	\mathcal{C}_{a,i}^{FFB}(x_0) = 0.468239957569145876297846606135,
	\\&
	\mathcal{C}_{a,i}^{FAB}(x_0) = -0.128836049327094055088259653772,
	\\&
	\mathcal{C}_{a,i}^{FBB}(x_0) = -0.0047313889291568923452785036443.
	\end{align}
\end{subequations}
The various color-structure components of ${\cal C}_{s}^{(2)}(x_0)$ and   ${\cal C}_s^{(3)}(x_0)$   read:
\begin{subequations}
\begin{align}
	&\mathcal{C}_{s}^{FF}(x_0) = -6.96020737354849312657205418357,
	\\
	&\mathcal{C}_{s}^{FA}(x_0) = -4.12970820397051570036738297443,
	\\
	&\mathcal{C}_{s}^{FB}(x_0) = 0.048170796075386686136602545271,
	\\
	&\mathcal{C}_{s}^{FC}(x_0) = 0.268781876466689639100436656736,
	\\
	&\mathcal{C}_{s}^{FL}(x_0) = -0.363661393326874053432216153222.
\\	&
 \mathcal{C}_{s}^{FFF}(x_0) = -12.6512824902497489841790999287,
	\\
	&  \mathcal{C}_{s}^{FFA}(x_0) = -91.3076763843495687930187876995,
	\\
	&  \mathcal{C}_{s}^{FAA}(x_0) = -67.2034246352357623358462321068,
	\\
	&  \mathcal{C}_{s}^{FFL}(x_0) = 31.12323218543900065296825277243,
	\\
	&  \mathcal{C}_{s}^{FAL}(x_0) = 19.49987491622541782889333507621,
	\\
	&  \mathcal{C}_{s}^{FCL}(x_0) = -0.262383795777090819586462489511,
	\\
	&  \mathcal{C}_{s}^{FBL}(x_0) = -0.0100054359359386271416783335057,
	\\
	&  \mathcal{C}_{s}^{FLL}(x_0) = 0.3393017746199103339986949104357,
	\\
	&  \mathcal{C}_{s}^{FBC}(x_0) = 0.0551829420149711792507691937024,
	\\
	&  \mathcal{C}_{s}^{FFC}(x_0) = 4.4105666464862568415402096694718,
	\\
	&  \mathcal{C}_{s}^{FAC}(x_0) = -0.6861454400278606762848799078670,
	\\
	&  \mathcal{C}_{s}^{FCC}(x_0) = 0.09536045103106728669303211267177,
	\\
	&  \mathcal{C}_{s}^{FFB}(x_0) = 1.1373781611175929139065523549900,
	\\
	&  \mathcal{C}_{s}^{FAB}(x_0) = -0.318775996588438248091557753877,
	\\
	&  \mathcal{C}_{s}^{FBB}(x_0) = 0.00994219408487397025982150577842.
	\end{align}
\end{subequations}
From the above numerical values, we find the dominant contributions in $\mathcal{C}_J^{(2)}(x_0)$ and $\mathcal{C}_J^{(3)}(x_0)$ come from the components corresponding to the color structures $C_F^2$, $C_FC_A$, $C_F^2C_A$ and $C_FC_A^2$, and the contributions from the bottom and charm quark loops are negligible.

Fixing the renormalization scale $\mu=\mu_0=3\mathrm{GeV}$, $m_b=4.75\mathrm{GeV}$, $m_c=1.5\mathrm{GeV}$, and setting the factorization scale $\mu_f=1.2\,\mathrm{GeV}$, Eq.~\eqref{Cjformula} then reduces to
\begin{align}\label{Cjasnum}
&\mathcal{C}_{v}
 =1-2.067273~\frac{\alpha_s^{(3)}(\mu_0)}{\pi}-29.29166\left(\frac{\alpha_s^{\left(3\right)}(\mu_0)}{\pi}\right)^2
 -1689.867\left(\frac{\alpha_s^{\left(3\right)}(\mu_0)}{\pi}\right)^3
 +\mathcal{O}(\alpha_s^4),
\nonumber\\
&\mathcal{C}_p
 =1-1.400607~\frac{\alpha_s^{\left(3\right)}(\mu_0)}{\pi}-27.80076\left(\frac{\alpha_s^{\left(3\right)}(\mu_0)}{\pi}\right)^2
 -1781.283\left(\frac{\alpha_s^{\left(3\right)}(\mu_0)}{\pi}\right)^3
 +\mathcal{O}(\alpha_s^4),
\nonumber\\
&\mathcal{C}_{a,i} =1-0.7339400~\frac{\alpha_s^{\left(3\right)}(\mu_0)}{\pi}-18.16765\left(\frac{\alpha_s^{\left(3\right)}(\mu_0)}{\pi}\right)^2
-922.7452\left(\frac{\alpha_s^{\left(3\right)}(\mu_0)}{\pi}\right)^3
+\mathcal{O}(\alpha_s^4),
\nonumber\\
&\mathcal{C}_s =1-0.06727332~\frac{\alpha_s^{\left(3\right)}(\mu_0)}{\pi}-15.46391\left(\frac{\alpha_s^{\left(3\right)}(\mu_0)}{\pi}\right)^2
-935.4686\left(\frac{\alpha_s^{\left(3\right)}(\mu_0)}{\pi}\right)^3
+\mathcal{O}(\alpha_s^4).
\end{align}

\begin{figure}[thb]
	\centering
	\includegraphics[width=0.44\textwidth]{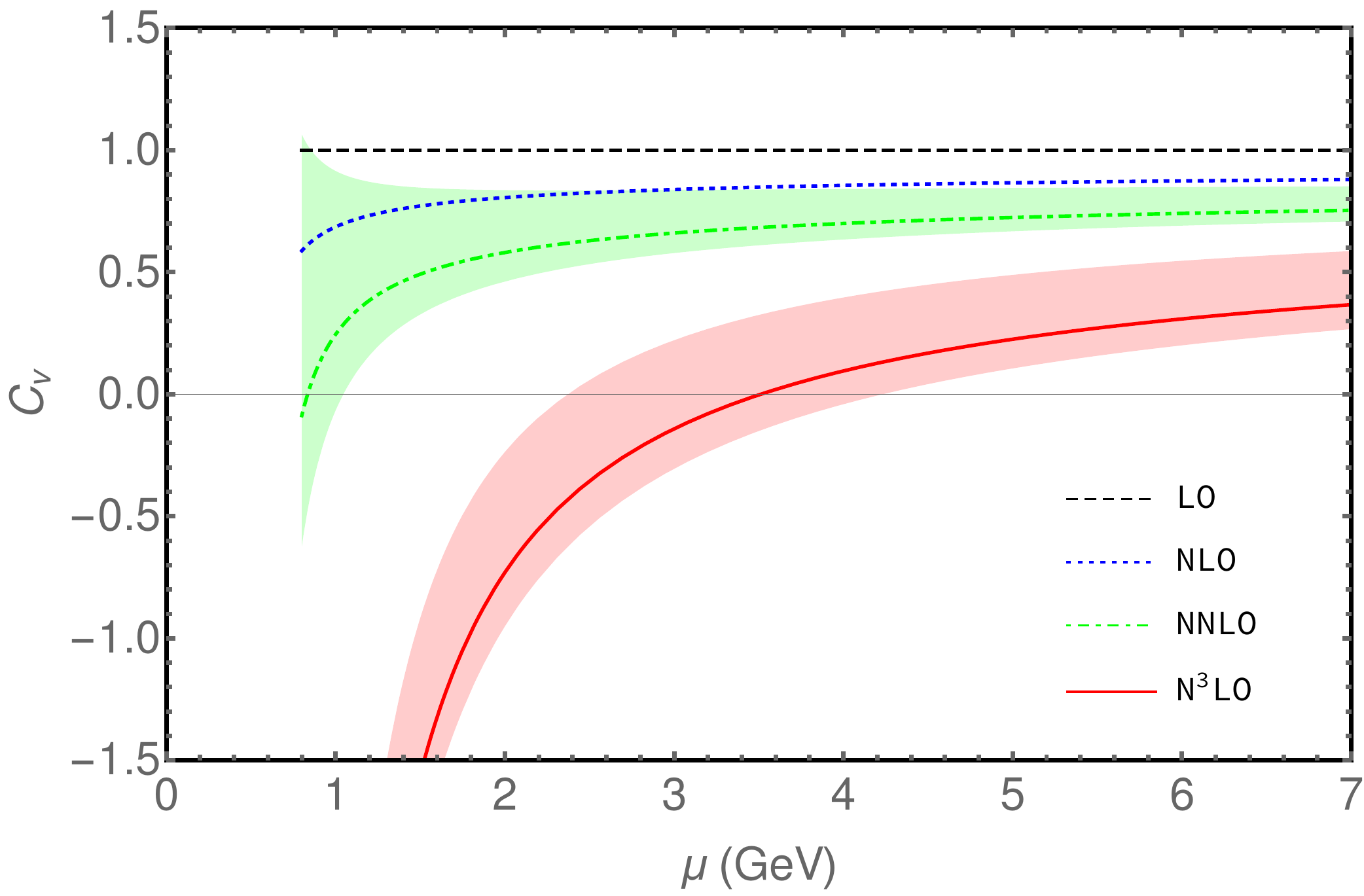}\qquad
	\includegraphics[width=0.44\textwidth]{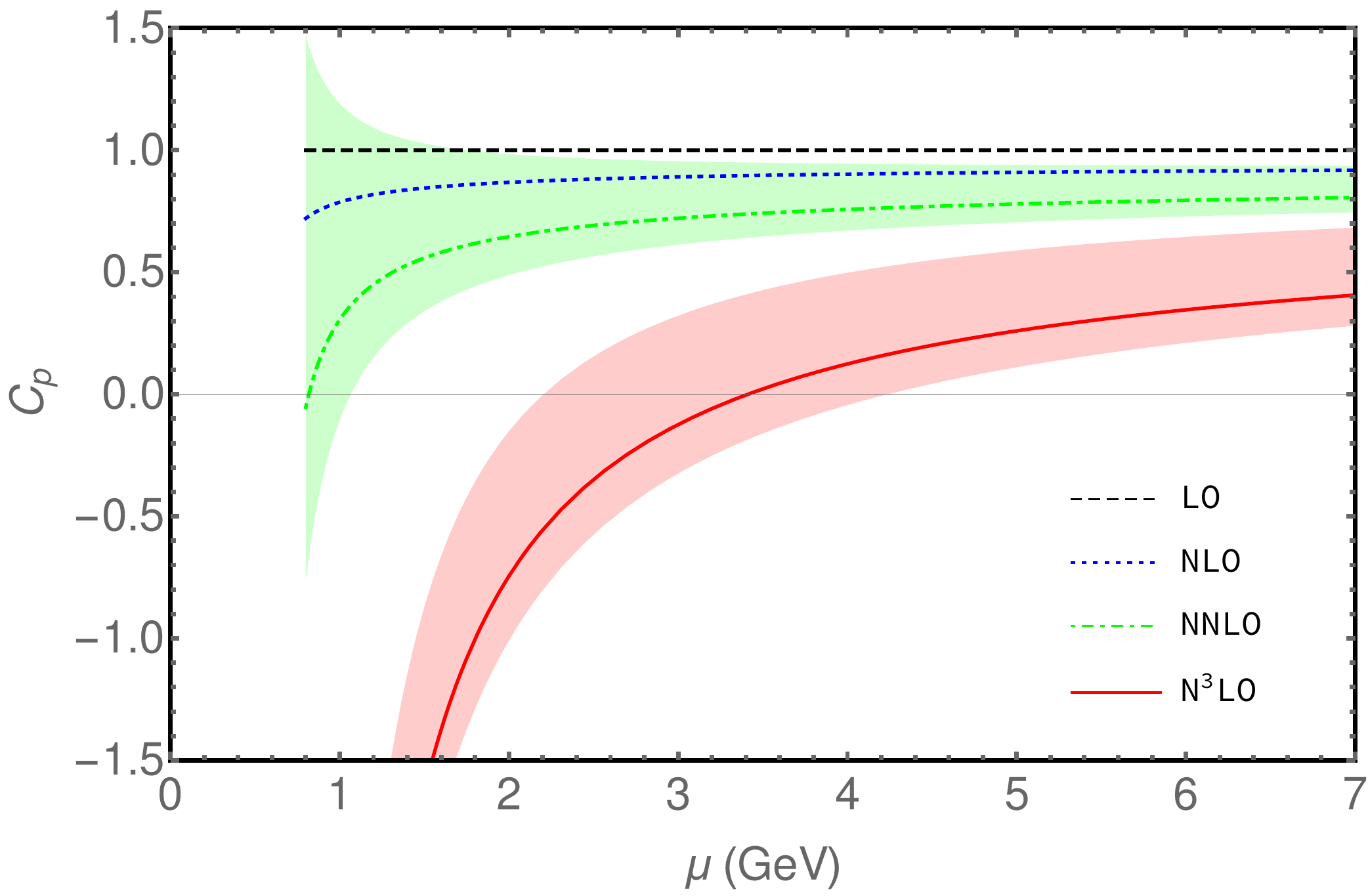}\qquad
	\includegraphics[width=0.44\textwidth]{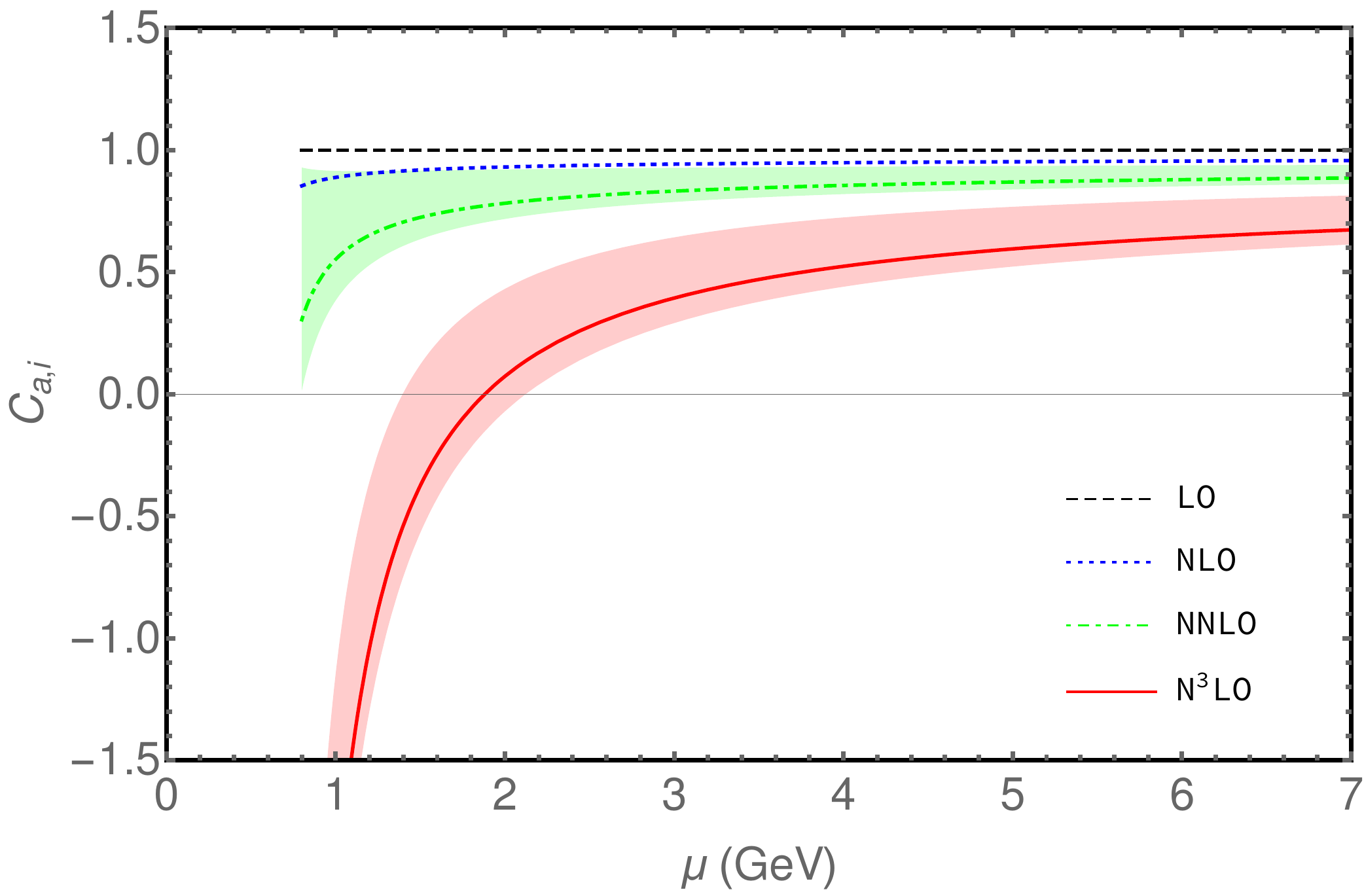}\qquad
	\includegraphics[width=0.44\textwidth]{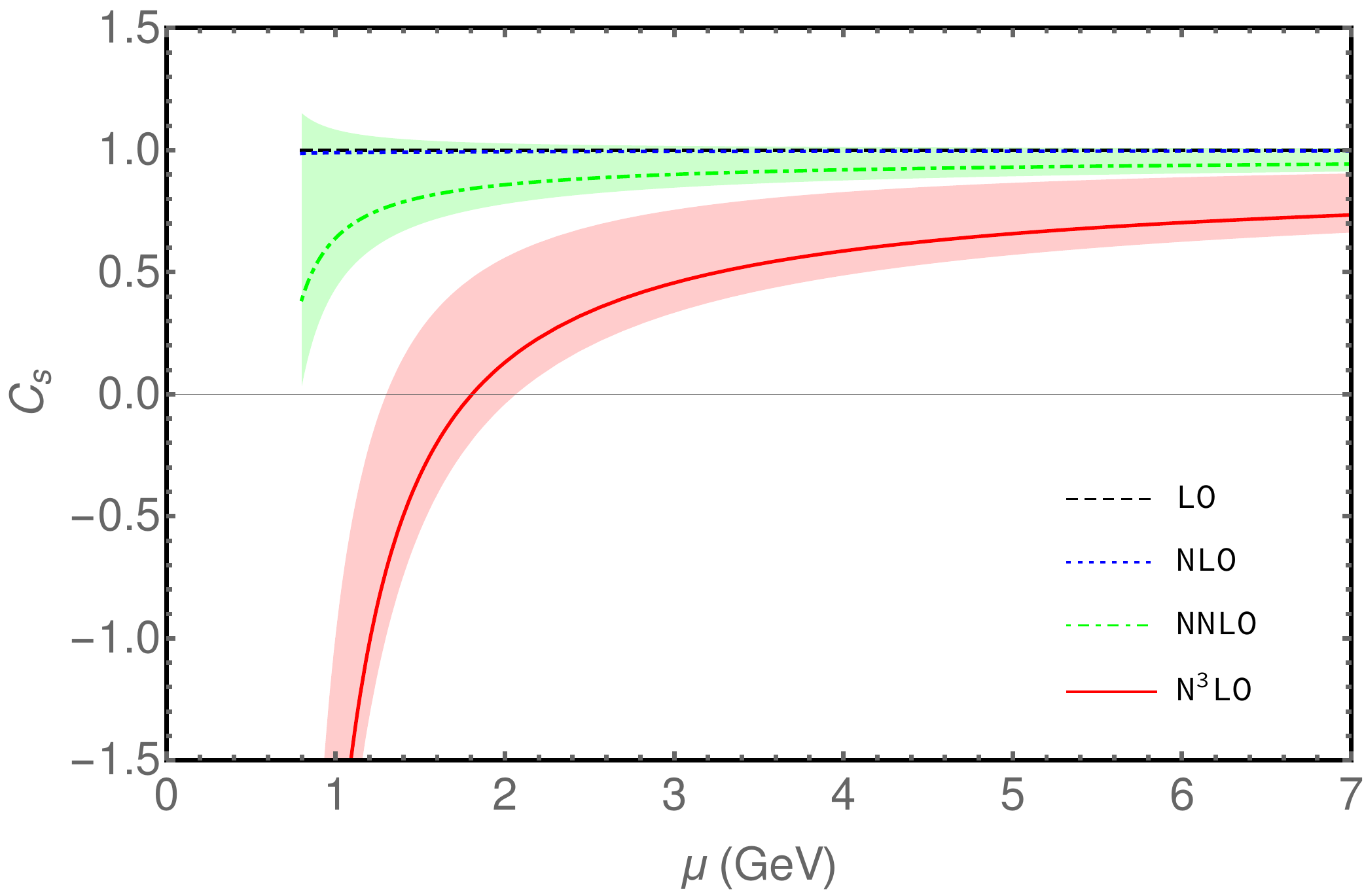}\qquad
	\caption{The renormalization scale $\mu$ dependence of the matching coefficients $\mathcal{C}_{J}$ for the heavy flavor-changing  currents
		at LO,  NLO,  NNLO and N$^3$LO accuracy. The central values of  the matching coefficients $\mathcal{C}_{J}$ are calculated inputting the  physical values with $\mu_f=1.2~\,\mathrm{GeV}$,  $m_b=4.75\mathrm{GeV}$ and $m_c=1.5\mathrm{GeV}$.   The error bands come from the variation of  $\mu_f$  between   2 and 0.4 $\mathrm{GeV}$. }
	\label{fig:Cjmudepend}
\end{figure}

With the values of $\alpha_s^{\left(n_l=3\right)}(\mu)$ calculated by the renormalization group running equation Eq.~\eqref{asrun2},
we investigate the renormalization scale dependence of the matching coefficients $\mathcal{C}_J$ for the heavy flavor-changing  currents
at LO,  NLO,  NNLO and N$^3$LO accuracy in Fig.~\ref{fig:Cjmudepend}.
The middle lines correspond to the choice of $\mu_f=1.2\,\mathrm{GeV}$ for the NRQCD factorization scale, and the upper and lower  edges   of the error bands
correspond to $\mu_f=0.4\,\mathrm{GeV}$ and $\mu_f=2~\mathrm{GeV}$, respectively.
We also present our precise numerical results of the matching coefficients $\mathcal{C}_{J}$  at LO,  NLO ,  NNLO and N$^3$LO in Table~\ref{tab:Cjnum}, where the uncertainties from $\mu_f$ and $\mu$ are included.
\begin{table}[thb]
	\begin{center}
		\caption{The values of the matching coefficients $\mathcal{C}_{J}$ for the heavy  flavor-changing   currents   up to N$^3$LO.
			The central values of  the matching coefficients $\mathcal{C}_{J}$ are calculated inputting the  physical values with $\mu_f=1.2~\mathrm{GeV}$, $\mu=\mu_0=3\mathrm{GeV}$, $m_b=4.75\mathrm{GeV}$ and $m_c=1.5\mathrm{GeV}$.
			The errors are estimated by varying $\mu_f$  from   2 to 0.4 $\mathrm{GeV}$, $\mu$  from   7 to 2.2 $\mathrm{GeV}$, respectively.}
		\label{tab:Cjnum}
		\renewcommand\arraystretch{2}
		\tabcolsep=0.3cm
		\begin{tabular}{ c c c c c}
			\hline\hline
			& LO         &  NLO                   & NNLO     & N$^3$LO
					\\  \hline
		$\mathcal{C}_{v}$	 & $1$ & $0.83875^{-0+0.04086}_{+0-0.02386}$   &    $0.66053^{-0.08198+0.09301}_{+0.17632-0.05738}$  &     $-0.14143^{-0.16360+0.50840}_{+0.36305-0.40923}$
		\\  \hline
		$\mathcal{C}_p$	 & $1$ & $0.89075^{-0+0.02768}_{+0-0.01617}$   &    $0.72161^{-0.10851+0.08443}_{+0.23336-0.05426}$  &     $-0.12374^{-0.20036+0.53021}_{+0.44615-0.43161}$	
						\\  \hline
			$\mathcal{C}_{a,i}$	 & $1$ & $0.94275^{-0+0.01451}_{+0-0.00847}$   &    $0.83222^{-0.04440+0.05394}_{+0.09549-0.03540}$  &     $0.39431^{-0.10240+0.27915}_{+0.24866-0.22332}$
			\\  \hline
			$\mathcal{C}_s$	 & $1$ & $0.99475^{-0+0.00133}_{+0-0.00078}$   &    $0.90067^{-0.05435+0.04210}_{+0.11688-0.02993}$  &     $0.45672^{-0.12257+0.27772}_{+0.29909-0.22662}$
			\\		\hline \hline
		\end{tabular}
	\end{center}
\end{table}
From Eq.~\eqref{Cjasnum}, Fig.~\ref{fig:Cjmudepend} and Table~\ref{tab:Cjnum}, one can find the higher order QCD corrections have larger values, especially, the ${\cal O}(\alpha_s^3)$ correction looks quite sizable, which confirms the nonconvergence behaviors of the matching coefficients investigated in previous literature~\cite{Egner:2022jot,Feng:2022ruy,Sang:2022tnh,Tao:2022hos}. From  Fig.~\ref{fig:Cjmudepend} and Table~\ref{tab:Cjnum}, it seems that both the NRQCD factorization scale dependence and the QCD renormalization scale dependence become larger at higher order.
Note, at each truncated perturbative order, the matching coefficient $\mathcal{C}_{J}$  is renormalization-group invariant~\cite{Feng:2022ruy,Sang:2022tnh}, e.g., at N$^3$LO, $\mathcal{C}_{J}$ obeys the following renormalization-group running invariance:
\begin{align}\label{runinvariance}
&\mathcal{C}_{J}^{\rm N^3LO}(\mu_f,\mu,m_b,m_c) =\mathcal{C}_{J}^{\rm N^3LO}(\mu_f,\mu_0,m_b,m_c) +\mathcal{O}(\alpha_s^4),
\end{align}
where $\mathcal{C}_{J}^{\rm N^3LO}(\mu_f,\mu,m_b,m_c)$ has dropped the $\mathcal{O}(\alpha_s^4)$ terms in Eq.~\eqref{Cjformula}. So the  $\mu$-dependence of the  N$^3$LO results is at the $\mathcal{O}(\alpha_s^4)$ order. Though the  $\mu$-dependence is suppressed by $\alpha_s^4$, the coefficients of $\alpha_s^4$  in  above equation  contain $\mu$-independent terms such as $\mathcal{C}_J^{(3)}(x)$ and $\ln \mu_f$, which come from the $\mathcal{O}(\alpha_s^3)$ order in Eq.~\eqref{Cjformula}, and have   considerably large values by aforementioned  calculation within the framework of NRQCD factorization. As a result, these terms will lead to a significantly larger renormalization scale dependence at N$^3$LO. From Fig.~\ref{fig:Cjmudepend}, we also find the NRQCD factorization scale $\mu_f$ has a dominant influence on the higher order QCD corrections.
When $\mu_f$ decreases, both the convergence of $\alpha_s$ expansion  and the independence of $\mu$ will improve.

Since  the  matching coefficients   on their own  are nonphysical, in order to obtain  a  reliable higher order correction to a physical quantity, the large $\alpha_s$-expansion nonconvergence  and the strong scale  dependence of the  matching coefficients at higher order  especially  $\mathcal{O}(\alpha_s^3)$     have to be  compensated by other higher order corrections.
Within the NRQCD effective  theroy,  physical quantities such as    the   beauty-charmed meson $B_{c}$ and $B^*_{c}$ decay constants   are factorized to the  matching coefficients   multiplied  with the long-distance nonperturbative NRQCD matrix elements (the Schr\"{o}dinger  wave functions at the origin), therefore we  not only need to calculate the  higher   order  QCD  corrections to the matching coefficients, but also need to perform the  higher   order corrections to the    wave functions at the origin.
Besides, it is also  indispensable to take into account  higher order relativistic corrections, resummation techniques and so on ( also see   Refs.~\cite{Beneke:2014qea,Marquard:2014pea,Czarnecki:1997vz,Beneke:1997jm,Egner:2022jot,Onishchenko:2003ui} for more discussions).

\section{Phenomenological analysis~\label{Phenomenological}}
Before starting our discussions about phenomenological applications of the matching coefficients, one can  review related  discussions in Refs.~\cite{Marquard:2014pea,Beneke:2014qea}. For  an explicit example, one can also see Eq.(4) in Ref.~\cite{Beneke:2014qea}, where the leptonic decay width of $\Upsilon$ meson was calculated up to N$^3$LO within NRQCD and pNRQCD effective theory and 
a large cancellation between three-loop perturbative QCD corrections to the matching coefficient and N$^3$LO perturbative corrections to the wave function at the origin was presented.
As following, we will confirm this phenomenon is also in the $c\bar{b}$ meson systems and   evaluate three-loop corrections to the decay constants, leptonic decay widths and corresponding branching ratios of the beauty-charmed mesons $B_{c}$ and $B^*_{c}$.

We use the following formulas ~\cite{Chung:2023mgr,Hwang:1999fc,Braaten:1995ej,Onishchenko:2003ui,Lee:2010ts,Beneke:1997jm,Penin:2014zaa,Beneke:2014qea,Feng:2022vvk} to compute
the decay constants  $f_{B_c^*}$ of the  vector  $B_c^*$ meson and $f_{B_c}$ of the pseudoscalar  $B_c$ meson:
\begin{align}
&f_{B_c^*}=2~\sqrt{\frac{N_c}{m_{B_c^*}}}~
\bigg[\mathcal{C}_{v}+\frac{d_v E_{B_c^*}}{12}\left(\frac{8}{M}-\frac{3}{m_r}\right)\bigg]
~|\Psi_{B_c^*}(0)|,\label{fBcv}
\\&
f_{B_c}=2~\sqrt{\frac{N_c}{m_{B_c}}}~
\bigg[\mathcal{C}_{p}-\frac{d_p E_{B_c}}{4m_r}\bigg]
~|\Psi_{B_c}(0)|,\label{fBc}
\end{align}
where $M=m_b+m_c$, $m_r=m_b m_c/(m_b+m_c)$ is the reduced mass, $m_{B_c^*}$ and $m_{B_c}$ are the masses of the vector and pseudoscalar  $c\bar{b}$ mesons, respectively.
Note that,  in our calculation we do not  employ the expansion  $m_{B_c^*}(m_{B_c})=m_b+m_c+E_{B_c^*}(E_{B_c})$ as Refs.~\cite{Beneke:2014qea,Chung:2023mgr,Beneke:2005hg} (in Table~\ref{tab:asseriesallv} and Table~\ref{tab:asseriesallp} we will see the influence from the expansion is small), but directly adopt the following physical values~\cite{Workman:2022ynf,Mathur:2018epb,Gomez-Rocha:2016cji,Zhou:2017svh,Yang:2021crs,Gregory:2009hq,Fulcher:1998ka,Ikhdair:2003ry,Martin-Gonzalez:2022qwd,Colquhoun:2015oha,Chaturvedi:2022pmn} for the $c\bar{b}$ mesons masses:
\begin{align}
&	m_{B_c}=6.274 \mathrm{GeV},~~ m_{B_c^*}=6.331 \mathrm{GeV}.
\end{align}

$d_v$ and $d_p$ are matching coefficients between QCD and NRQCD in sub-leading order of relative velocity for the vector  $B_c^*$ meson and pseudoscalar  $B_c$ meson, respectively.
One-loop results of $d_v$ and $d_p$ can be found in Refs.~\cite{Hwang:1999fc,Lee:2010ts,Beneke:2007gj}
\begin{align}
&d_v=1-\frac{\alpha_s^{(n_l)}(\mu)}{\pi} \frac{C_F(x+1)^2}{3 x^2-2 x+3} \left(\frac{4 }{3 }+\frac{(x-1) \left(7
	x^2+6 x+7\right) }{4  (x+1)^3}\ln x+4 \ln \frac{\mu_f ^2}{m_b m_c}\right)+\mathcal{O}(\alpha_s^2),\label{dv}
\\&
d_p=1+\frac{\alpha_s^{(n_l)}(\mu)}{\pi} C_F \left(\frac{x^2+98 x+1}{18 (x+1)^2}-\frac{(x-1) \left(7 x^2+46 x+7\right) }{12
	(x+1)^3}\ln x-\frac{4}{3} \ln \frac{\mu_f ^2}{m_b m_c}\right)+\mathcal{O}(\alpha_s^2).\label{dp}
\end{align}

$E_{B_c^*}$ and   $|\Psi_{B_c^*}(0)|$ are  the binding energy and the wave function  at the origin for $B_c^*$, respectively, while $E_{B_c}$ and $|\Psi_{B_c}(0)|$ are for $B_c$.   
In the previous sections, we have calculated higher-order perturbative corrections to the matching coefficient by extracting the hard contribution using the NRQCD effective theory. Similarly, the  higher-order perturbative corrections to the binding energy and the wave function  at the origin can be calculated allowing for the soft, potential and ultrasoft contributions from the Coulomb potential, non-Coulomb potentials and ultrasoft gluon exchange     using the potential NRQCD effective theory  (pNRQCD)~\cite{Pineda:1997bj,Beneke:1999zr,Brambilla:1999xf,Luke:1999kz,Brambilla:2004jw}.
Since the pseudoscalar meson $B_c({1}^1{S_0})$ and vector meson $B_c^*({1}^3{S_1})$ are respectively  the  lowest-lying  spin-singlet and   spin-triplet $S$-wave bound state, i.e., $B_c$ and $B_c^*$ have different spin quantum  numbers, $E_{B_c}$ and     $|\Psi_{B_c}(0)|$ are slightly different (see Refs.~\cite{Beneke:2005hg,Godfrey:2004ya,Penin:2005eu,Hwang:1997ie,Rai:2008sc,Kniehl:2002yv,Marquard:2006qi,Chung:2020zqc}) from $E_{B_c^*}$ and $|\Psi_{B_c^*}(0)|$, respectively. 
However, the binding energies of the double heavy quark systems are spin-independent  up to NLO (see Refs.~\cite{Penin:2005eu,Beneke:2005hg,Schuller:2008rxa}), hence $E_{B_c}^{\text{NLO}}\equiv E_{B_c^*}^{\text{NLO}}$, which 
can be obtained from Refs.~\cite{Peset:2015vvi,Brambilla:2000db} as
\begin{align}\label{ebc}
E_{B_c}^{\text{NLO}}\equiv E_{B_c^*}^{\text{NLO}}= E^{(0)}\Bigg\{1+\frac{\alpha_s^{(n_l)}(\mu_f) }{\pi }\Bigg[\beta _0^{(n_l)}\left(\ln\left( \frac{\mu _f}{2 m_r C_F \alpha_s^{(n_l)}(\mu_f)}\right)+\frac{11}{6}\right)-\frac{4}{3} C_A \Bigg]\Bigg\},
\end{align} 
where  $E^{(0)}=-\frac{m_r}{2}\left(\alpha_s^{(n_l)}(\mu_f)C_F\right)^2 $,  $n_l=3$. For simplicity, we have set the pNRQCD factorization scale equal to the NRQCD factorization scale $\mu_f$~\footnote{ This is not a problem because, in strict fixed order computations it is
not necessary to distinguish between factorization scales as they all cancel to the required accuracy  once all contributions to an observable are added~\cite{Pineda:2011dg}.},  which separates contributions coming from the hard and soft momentum regions.   $\alpha_s^{(n_l)}(\mu_f)$ can be translated into $\alpha_s^{(n_l)}(\mu)$  by the coupling running equation in Eq.\eqref{asrun1}.
Based on Eq.~\eqref{fBcv} and Eq.~\eqref{fBc}, the NLO results in Eq.~\eqref{dv}, Eq.~\eqref{dp} and Eq.~\eqref{ebc} are    sufficient  for the N$^3$LO calculation of the decay constants.

For phenomenological analysis of the pseudoscalar $B_c$ meson, we will use the approximation  $|\Psi_{B_c}(0)|\approx |\Psi_{B_c^*}(0)|$, so that the only  missing piece  in Eq.~\eqref{fBcv} and Eq.~\eqref{fBc} is $|\Psi_{B_c^*}(0)|$. 
In order to obtain $|\Psi_{B_c^*}(0)|$, we  employ the scale relation explored  by Collins $et~al.$~\cite{Collins:1996pp,Fulcher:1998ka,Patel:2008na}:
\begin{align}\label{bbcc2bc}
|\Psi_{B_c^*}(0)|=|\Psi_{J/\psi}(0)|^{1-y}~|\Psi_{\Upsilon}(0)|^{y},
\end{align}
where  $y\in[0.3,0.4]$ with the central value $y=y_c=\ln((1+m_c/m_b)/2)/\ln(m_c/m_b)$~\footnote{
	The central value $y=y_c=\ln((1+m_c/m_b)/2)/\ln(m_c/m_b)$  can be exactly derived from the scale law of the wave function at the origin: $|\psi_{\mu}^n(0)|^2=f(n,a)(A\mu)^{3/(2+a)}$ for any reduced mass $\mu$ and any class of power-law potentials $V(r)=A r^a+C$, where $f(n,a)$ is only a function of the radial quantum number $n$ and the power $a$. 
	To cover the uncertainties from various potentials, the maximal (minimal) value of $y$ is simply taken to be 0.4 (0.3) independent of the quark masses. 
For more details, see Ref.~\cite{Collins:1996pp} (where $y=0.35$ was chosen as the only optimal solution) .} is the scale power. $|\Psi_{J/\psi}(0)|$ and $|\Psi_{\Upsilon}(0)|$ are the wave functions at the origin for the vector heavy quarkonia  $J/\psi$ and   $\Upsilon$, respectively.
From literature~\cite{Beneke:2007uf,Beneke:2013jia,Beneke:2014qea,Peset:2015vvi,Shen:2015cta,Schroder:1998vy,Schuller:2008rxa,Beneke:2005hg,Beneke:2007gj,Beneke:2007pj,Beneke:2016kkb,Penin:2014zaa,Beneke:2014pta,Chung:2023mgr,Beneke:1999zr},  the full N$^3$LO  corrections to the wave function at the origin for the lowest-lying vector heavy quarkonium $Q\bar{Q}$  can be expressed as 
\begin{align}\label{wavefun1}
\frac{|\Psi_{Q\bar{Q}}^V(0)|^2}{|\Psi_{Q\bar{Q}}^{(0)}(0)|^2}=&1+\frac{\alpha_s^{(n_f)}(\mu_f) }{\pi }\frac{6 \beta^{(n_f)}_0 L_s+c_{\psi,1}^{\text C}}{4} 
+\left(\frac{\alpha_s^{(n_f)}(\mu_f) }{\pi }\right)^2 \Bigg\{\frac{3}{2} {\beta _0^{(n_f)}}^2 L_s^2
\nonumber\\ &
+\frac{L_s}{24}  \big[9 {\beta _1^{(n_f)}}-18 {\beta_0^{(n_f)}}^2+12 {\beta _0^{(n_f)}}
c_{\psi,1}^{\text C}+8 \pi ^2 C_F \left(3 C_A+2 C_F\right)\big]
\nonumber\\ &
+\frac{1}{144} \big[2 \pi ^2 C_F \left(162 C_A+89 C_F\right)+9 c_{\psi,2}^{\text C}\big]
\Bigg\}
\nonumber\\ &
+
\left(\frac{\alpha_s^{(n_f)}(\mu_f) }{\pi }\right)^3\pi ^2 \Bigg\{
C_A^3 \bigg[-\frac{L_s}{4}+\frac{L_m}{4}-\ln(C_F)-\frac{\pi ^2}{12}+\frac{1}{2}\bigg]
\nonumber\\ &
+C_A^2 C_F \bigg[
-\frac{2}{3}	L_s^2+L_s \left(\frac{4}{3} L_m+4\ln 2+\frac{17}{18}\right)-\frac{L_m^2}{6}
\nonumber\\ &
+ L_m\left(\frac{31}{9}-\ln 2\right)
+\frac{2}{9} \ln \left(C_F\right) \left(9 \ln
\left(C_F\right)+12 \ln 2-44\right)\bigg]
\nonumber\\ &
+C_A C_F^2 \bigg[-\frac{9}{4}L_s^2 +L_s \left(\frac{9 }{2}L_m+4\ln
2-\frac{25}{54}\right)-\frac{3 }{8}L_m^2
\nonumber\\ &
+ L_m\left(\frac{61}{9}-\ln  2\right)
+\frac{4}{3} \ln \left(C_F\right) \left(4 \ln \left(C_F\right)+\ln 2-10\right)\bigg]
\nonumber\\ &
+
C_A C_F \bigg[2 {\beta_0^{(n_f)}} L_s^2+\frac{ L_s}{72} \left({\beta_0^{(n_f)}}\left(423-12 \pi ^2\right) -218 T_F n_f \right)\bigg]
\nonumber\\ &
+C_F^3 \bigg[-\frac{3}{2} L_s^2+L_s \left(3 L_m-8 \ln
2+\frac{47}{18}\right)-\frac{L_m^2}{4}
\nonumber\\ &
+L_m\left(2\ln 2+\frac{59}{36}\right) 
+\frac{8}{9} \ln \left(C_F\right) \left(3 \log \left(C_F\right)-9 \ln 2+2\right)\bigg]
\nonumber\\ &
+C_F^2 T_F \bigg[L_s\left(\frac{2}{15}-\frac{59}{27} n_f\right)
+\frac{L_m}{15}\bigg]
+{\beta_0^{(n_f)}} C_F^2 \bigg[\frac{4 }{3}L_s^2+L_s\left(\frac{53}{16}-\frac{\pi
	^2}{9}\right) \bigg]
\nonumber\\ &
+\frac{1}{\pi ^2}\bigg[
\frac{5}{4}
{\beta_0^{(n_f)}}^3 L_s^3
+\frac{{\beta_0^{(n_f)}}L_s^2}{32}  \left(27 {\beta_1^{(n_f)}}-54 {\beta_0^{(n_f)}}^2+20 {\beta_0^{(n_f)}} c_{\psi,1}^{\text C}\right) 
\nonumber\\ &
+\frac{ L_s}{32} \left(12 {\beta_0^{(n_f)}}^3-12 {\beta_1^{(n_f)}} {\beta_0^{(n_f)}}+3 {\beta_2^{(n_f)}}-8 {\beta_0^{(n_f)}}^2 c_{\psi,1}^{\text C}
\right.\nonumber\\ &\left.
+4 {\beta_1^{(n_f)}} c_{\psi,1}^{\text C}+5 {\beta_0^{(n_f)}}
c_{\psi,2}^{\text C}\right)
+\frac{c_{\psi,3}^{\text C}+c_{\psi,3}^{\text{nC}}}{64}
\bigg]
+\delta_1^{\text{us}}\Bigg\}+\mathcal{O}(\alpha_s^4),
\end{align}
where $|\Psi_{Q\bar{Q}}^{(0)}(0)|^2=(m\alpha_s^{(n_f)}(\mu_f)C_F)^3/(8\pi)$,  $L_s=\ln[\mu_f/(m\alpha_s^{(n_f)}(\mu_f)C_F)]$, $L_m=\ln(\mu_f/m)$, $m$ is the mass of $Q$, $n_f$ is  flavours   lighter than  $Q$, and the non-logarithmic terms of the Coulomb, non-Coulomb and ultrasoft corrections read
\begin{align}
c_{\psi,1}^{\text C} &= 2.62290 - 1.61351 n_f,
\nonumber\\
c_{\psi,2}^{\text C} &= 1800.745 - 193.4887 n_f + 3.50376 n_f^2,
\nonumber\\
c_{\psi,3}^{\text C} &= -39854.2 + 2005.08 n_f + 19.79845 n_f^2 + 3.61806 n_f^3,
\nonumber\\
c_{\psi,3}^{\text {nC}} &= -44754.7 - 3126.52 n_f,
\nonumber\\
\delta_1^{\text{us}} &= 353.06.
\end{align} 
Equivalently, Eq.~\eqref{wavefun1} can  be simplified in a numerical form
\begin{align}
\frac{|\Psi_{Q\bar{Q}}^V(0)|^2}{|\Psi_{Q\bar{Q}}^{(0)}(0)|^2}=&1+
\alpha_s^{(n_f)}(\mu_f) \bigg[L_s \left(5.252113122\, -0.3183098862 n_f\right)-0.1283989789 n_f
\nonumber\\&
+0.2087238344\bigg]
+\left(\alpha_s^{(n_f)}(\mu_f)\right)^2 \bigg[L_s^2\left(0.06754745576 n_f^2-2.229066040 n_f
\right.\nonumber\\&\left.
+18.38979483\right)
+ L_s\left(0.02072049126 n_f^2-0.3544823646 n_f+1.327477963\right)
\nonumber\\&
+0.02218780425 n_f^2-1.225281344 n_f+22.60088339\bigg]
\nonumber\\&
+\left(\alpha_s^{(n_f)}(\mu_f)\right)^3 \bigg[L_s^3\left(-0.01194501275 n_f^3+0.5912781313 n_f^2
-9.756089166 n_f
\right.\nonumber\\&\left.
+53.65849041\right)
+L_s^2\left(0.001670726641 n_f^3
-0.06791901544 n_f^2
\right.\nonumber\\&\left.
+0.6172398127 n_f-6.696584076\right)
+L_s \left(14.99593242
L_m
\right.\nonumber\\&\left.
-0.009572483336 n_f^3+0.8198924520 n_f^2
-23.92535721 n_f
+192.0183234\right)
\nonumber\\&
+0.001823251079 n_f^3+0.009977039176 n_f^2
-0.5651253308 n_f
\nonumber\\&
+50.45825046
-1.461867625 L_m^2+25.28876373 L_m
\bigg]+\mathcal{O}(\alpha_s^4).
\end{align}
With $|\Psi_{Q\bar{Q}}^V(0)|$ known, we can obtain  $|\Psi_{J/\psi}(0)|$ and $|\Psi_{\Upsilon}(0)|$  as following
\begin{align}
\Psi_{J/\psi}(0)&=\Psi_{Q\bar{Q}}^V(0)|_{m\to m_c,n_f\to n_l},
\nonumber\\
\Psi_{\Upsilon}(0)&=\Psi_{Q\bar{Q}}^V(0)|_{m\to m_b,n_f\to n_l+1},
\end{align}
where $n_l=3$ denotes the massless flavours. Furthermore, we  use the decoupling relation in Eq.~\eqref{Decoupling} (but up to $\mathcal{O}(\alpha_s^3)$) to translate $\alpha_s^{(n_l+1)}(\mu_f)$ into $\alpha_s^{(n_l)}(\mu_f)$, which can further be translated into $\alpha_s^{(n_l)}(\mu)$ by the coupling running equation in Eq.~\eqref{asrun1} (but up to $\mathcal{O}(\alpha_s^3)$).
By Eq.~\eqref{bbcc2bc}, we can finally express $|\Psi_{B_c^*}(0)|$ in power series of $\alpha_s^{(n_l=3)}(\mu)$ and we have checked the obtained series expansion in $\alpha_s^{(n_l=3)}(\mu)$ of $|\Psi_{B_c^*}(0)|$ obeys the  renormalization-group running invariance as Eq.~\eqref{runinvariance}.

As following, we will present the numeric results for the matching coefficients $d_v$ and $d_p$ for sub-leading order relativistic corrections, the binding energies, the wave functions at the origin, the decay constants, leptonic decay widths and corresponding branching ratios for the pseudoscalar and vector $c\bar{b}$ mesons.
In spirit of perturbation theory, through all our calculations we expand all of above quantities  in power series of $\alpha_s^{(n_l=3)}(\mu)$ by the decoupling relation with coupling running, and truncate them up to a fixed order. 
Choosing  the scale power $y=y_c$ and  fixing  the factorization scale $\mu_f=1.2\,\mathrm{GeV}$, the renormalization scale $\mu=\mu_0=3\,\mathrm{GeV}$, $m_b=4.75\mathrm{GeV}$, $m_c=1.5\mathrm{GeV}$, we present the  $\alpha_s$-expansion  of   these quantities in Table~\ref{tab:asseriesallv} and Table~\ref{tab:asseriesallp}, where the leading-order of the wave functions at the origin and the decay constants are determined by Eq.~\eqref{fBcv}, Eq.\eqref{fBc}  and Eq.\eqref{bbcc2bc} as:  
\begin{align}
|\Psi^{(0)}_{B_c}(0)|^2=|\Psi^{(0)}_{B_c^*}(0)|^2&=\frac{(m_r\alpha_s^{\left(n_l=3\right)}(\mu)C_F)^3}{\pi},
\nonumber\\
f_{B_c^*}^{(0)}&=2\sqrt{\frac{N_c}{m_{B_c^*}}}|\Psi^{(0)}_{B_c^*}(0)|,
\nonumber\\
f_{B_c}^{(0)}&=2\sqrt{\frac{N_c}{m_{B_c}}}|\Psi^{(0)}_{B_c}(0)|.
\end{align} 
From  Table~\ref{tab:asseriesallv} and Table~\ref{tab:asseriesallp}, we find  the influences of $d_v$, $d_p$, $E_{B_c^*}$  and $E_{B_c}$  on the decay constants in Eq.~\eqref{fBcv} and Eq.~\eqref{fBc} are inconsiderable compared  with the matching coefficients  $\mathcal{C}_{v}$ and $\mathcal{C}_{p}$. Both $\mathcal{C}_{v}$ ($\mathcal{C}_{p}$) and $|\Psi_{B_c^*}(0)|$ ($|\Psi_{B_c}(0)|$) have a large  $\alpha_s$-expansion nonconvergence at the third order, however, after a large cancellation between the matching coefficient  and the wave function at the origin, the perturbative $\alpha_s$-expansion of the  decay constant becomes  convergent  up to N$^3$LO.

\begin{table}[htbp]\tiny
	\caption{The expansion coefficients of $\left({\alpha_s^{\left(n_l=3\right)}(\mu_0)}/{\pi}\right)^i(i=0,1,2,3)$ with $y=y_c$, $\mu_f=1.2\,\mathrm{GeV}$, $\mu=\mu_0=3\,\mathrm{GeV}$, $m_b=4.75\mathrm{GeV}$, $m_c=1.5\mathrm{GeV}$ for the matching coefficients, the binding energies, the wave functions at the origin, the decay constants and the squared decay constants of the vector $B_c^*$ meson. For details see the text.}
	\label{tab:asseriesallv}
	\setlength{\tabcolsep}{0.3mm}
	\centering
	\renewcommand{\arraystretch}{1.5}
	{
		\begin{tabular}{c|c|c|c|c}
			\hline\hline
			{$$}
			& $\left(\frac{\alpha_s^{\left(3\right)}(\mu_0)}{\pi}\right)^0$
			& $\left(\frac{\alpha_s^{\left(3\right)}(\mu_0)}{\pi}\right)^1$
			& $\left(\frac{\alpha_s^{\left(3\right)}(\mu_0)}{\pi}\right)^2$ 
			& $\left(\frac{\alpha_s^{\left(3\right)}(\mu_0)}{\pi}\right)^3$   \\
			\hline
			\multirow{1}*{$\mathcal{C}_{v}$}
			& \multirow{1}*{$1$}
			& \multirow{1}*{$-2.0673$}
			& \multirow{1}*{$-29.292$}
			& \multirow{1}*{$-1689.9$}               \\
			\hline
			{$\frac{\mathcal{C}_{v} E_{B_c^*}}{M}$}
			& {$0$}
			& {$0$}
			&{$-1.6002$}
			& {      $-16.504 + 
				14.402 \ln\alpha_s^{\left(3\right)}(\mu_0)= -36.757$     }              \\
			\hline
			{$\frac{d_{v} E_{B_c^*}}{12}\left(\frac{8}{M}-\frac{3}{m_r}\right)$}
			& {$0$}
			& {$0$}
			&{$1.1265$}
			&   {    $18.072 - 
				10.138\ln\alpha_s^{\left(3\right)}(\mu_0)= 32.329$  }                 \\
			\hline
			\multirow{2}*{$\frac{|\Psi_{B_c^*}(0)|}{|\Psi_{B_c^*}^{(0)}(0)|}$}
			& \multirow{2}*{$1$}
			& $- 6.5684 \ln\alpha_s^{\left(3\right)}(\mu_0)$
			& $72.534 - 6.5326 \ln\alpha_s^{\left(3\right)}(\mu_0) $
			&        $- 183.97 \ln^3\alpha_s^{\left(3\right)}(\mu_0)- 
			74.826\ln^2\alpha_s^{\left(3\right)}(\mu_0) $                   \\
			&
			&$-0.17827=9.0589$
			& $+ 35.972 \ln^2\alpha_s^{\left(3\right)}(\mu_0)=152.86$
			&$ - 1231.0\ln\alpha_s^{\left(3\right)}(\mu_0) +639.26=2734.1$ \\
			\hline
			\multirow{2}*{$\frac{f_{B_c^*}}{f_{B_c^*}^{(0)}}$}
			& \multirow{2}*{$1$}
			& $ - 6.5684 \ln\alpha_s^{\left(3\right)}(\mu_0)$
			& $44.738 + 7.0460 \ln\alpha_s^{\left(3\right)}(\mu_0) $
			&        $ - 183.97 \ln^3\alpha_s^{\left(3\right)}(\mu_0) - 
			149.19 \ln^2\alpha_s^{\left(3\right)}(\mu_0)$                   \\
			&
			&$-2.2455=6.9916 $
			& $+ 
			35.972 \ln^2\alpha_s^{\left(3\right)}(\mu_0)=105.97$
			&$- 1042.7 \ln\alpha_s^{\left(3\right)}(\mu_0)-1177.5 =505.43$ \\
			\hline
			\multirow{2}*{$\frac{f_{B_c^*}^2}{{f_{B_c^*}^{(0)}}^2}$}
			& \multirow{2}*{$1$}
			& $ - 13.137 \ln\alpha_s^{\left(3\right)}(\mu_0)$
			& $94.518 + 43.591 \ln\alpha_s^{\left(3\right)}(\mu_0) $
			&        $ - 840.50 \ln^3\alpha_s^{\left(3\right)}(\mu_0) - 
			552.50 \ln^2\alpha_s^{\left(3\right)}(\mu_0)$                   \\
			&
			&$-4.4911 =13.983  $
			& $+ 	115.09\ln^2\alpha_s^{\left(3\right)}(\mu_0)=260.82$
			&$ - 2704.7 \ln\alpha_s^{\left(3\right)}(\mu_0)-2555.8 =2492.7$ \\
			\hline\hline
			\end{tabular}
		}
\end{table}

\begin{table}[htbp]\tiny
			\caption{The same as Table~\ref{tab:asseriesallv}, but for the pseudoscalar $B_c$ meson.}
			\label{tab:asseriesallp}
			\setlength{\tabcolsep}{0.3mm}
			\centering
			\renewcommand{\arraystretch}{1.5}
			{
				\begin{tabular}{c|c|c|c|c}
					\hline\hline
					{$$}
					& $\left(\frac{\alpha_s^{\left(3\right)}(\mu_0)}{\pi}\right)^0$
					& $\left(\frac{\alpha_s^{\left(3\right)}(\mu_0)}{\pi}\right)^1$
					& $\left(\frac{\alpha_s^{\left(3\right)}(\mu_0)}{\pi}\right)^2$ 
					& $\left(\frac{\alpha_s^{\left(3\right)}(\mu_0)}{\pi}\right)^3$   \\
					\hline
					\multirow{1}*{$\mathcal{C}_{p}$}
					& \multirow{1}*{$1$}
					& \multirow{1}*{$-1.4006$}
					& \multirow{1}*{$-27.801$}
					& \multirow{1}*{$-1781.3$}               \\
					\hline
					{$\frac{\mathcal{C}_{p} E_{B_c}}{M}$}
					& {$0$}
					& {$0$}
					&{$-1.6002$}
					& {      $-17.570 + 14.402\ln\alpha_s^{\left(3\right)}(\mu_0)= -37.824$     }              \\
					\hline
					{$-\frac{d_{p} E_{B_c}}{4 m_r}$}
					& {$0$}
					& {$0$}
					&{$2.1932$}
					&   {    $34.521 - 19.739\ln\alpha_s^{\left(3\right)}(\mu_0)= 62.280$  }                 \\
					\hline
					\multirow{2}*{$\frac{|\Psi_{B_c}(0)|}{|\Psi_{B_c}^{(0)}(0)|}$}
					& \multirow{2}*{$1$}
					& $- 6.5684 \ln\alpha_s^{\left(3\right)}(\mu_0)$
					& $72.534 - 6.5326 \ln\alpha_s^{\left(3\right)}(\mu_0) $
					&        $- 183.97 \ln^3\alpha_s^{\left(3\right)}(\mu_0)- 
					74.826\ln^2\alpha_s^{\left(3\right)}(\mu_0) $                   \\
					&
					&$-0.17827=9.0589$
					& $+ 35.972 \ln^2\alpha_s^{\left(3\right)}(\mu_0)=152.86$
					&$ - 1231.0\ln\alpha_s^{\left(3\right)}(\mu_0) +639.26=2734.1$ \\
					\hline
					\multirow{2}*{$\frac{f_{B_c}}{f^{(0)}_{B_c}}$}
					& \multirow{2}*{$1$}
					& $ - 6.5684 \ln\alpha_s^{\left(3\right)}(\mu_0)$
					& $47.177 + 2.6671 \ln\alpha_s^{\left(3\right)}(\mu_0) $
					&        $ - 183.97 \ln^3\alpha_s^{\left(3\right)}(\mu_0) - 
					125.21 \ln^2\alpha_s^{\left(3\right)}(\mu_0)$                   \\
					&
					&$-1.5789=7.6582 $
					& $+ 
					35.972 \ln^2\alpha_s^{\left(3\right)}(\mu_0)=114.57$
					&$- 1073.4 \ln\alpha_s^{\left(3\right)}(\mu_0)-1204.5 =569.05$ \\
					\hline
					\multirow{2}*{$\frac{f_{B_c}^2}{{f^{(0)}_{B_c}}^2}$}
					& \multirow{2}*{$1$}
					& $ - 13.137 \ln\alpha_s^{\left(3\right)}(\mu_0)$
					& $96.846 + 26.075 \ln\alpha_s^{\left(3\right)}(\mu_0) $
					&        $ - 840.50 \ln^3\alpha_s^{\left(3\right)}(\mu_0) - 
					399.05  \ln^2\alpha_s^{\left(3\right)}(\mu_0)$                   \\
					&
					&$-3.1577 =15.316 $
					& $+ 	115.09\ln^2\alpha_s^{\left(3\right)}(\mu_0)=287.78$
					&$ - 2775.0 \ln\alpha_s^{\left(3\right)}(\mu_0)-2558.0 =2892.9$ \\
					\hline\hline
					\end{tabular}
				}
\end{table}

\begin{figure}[thb]
	\centering
	\includegraphics[width=0.44\textwidth]{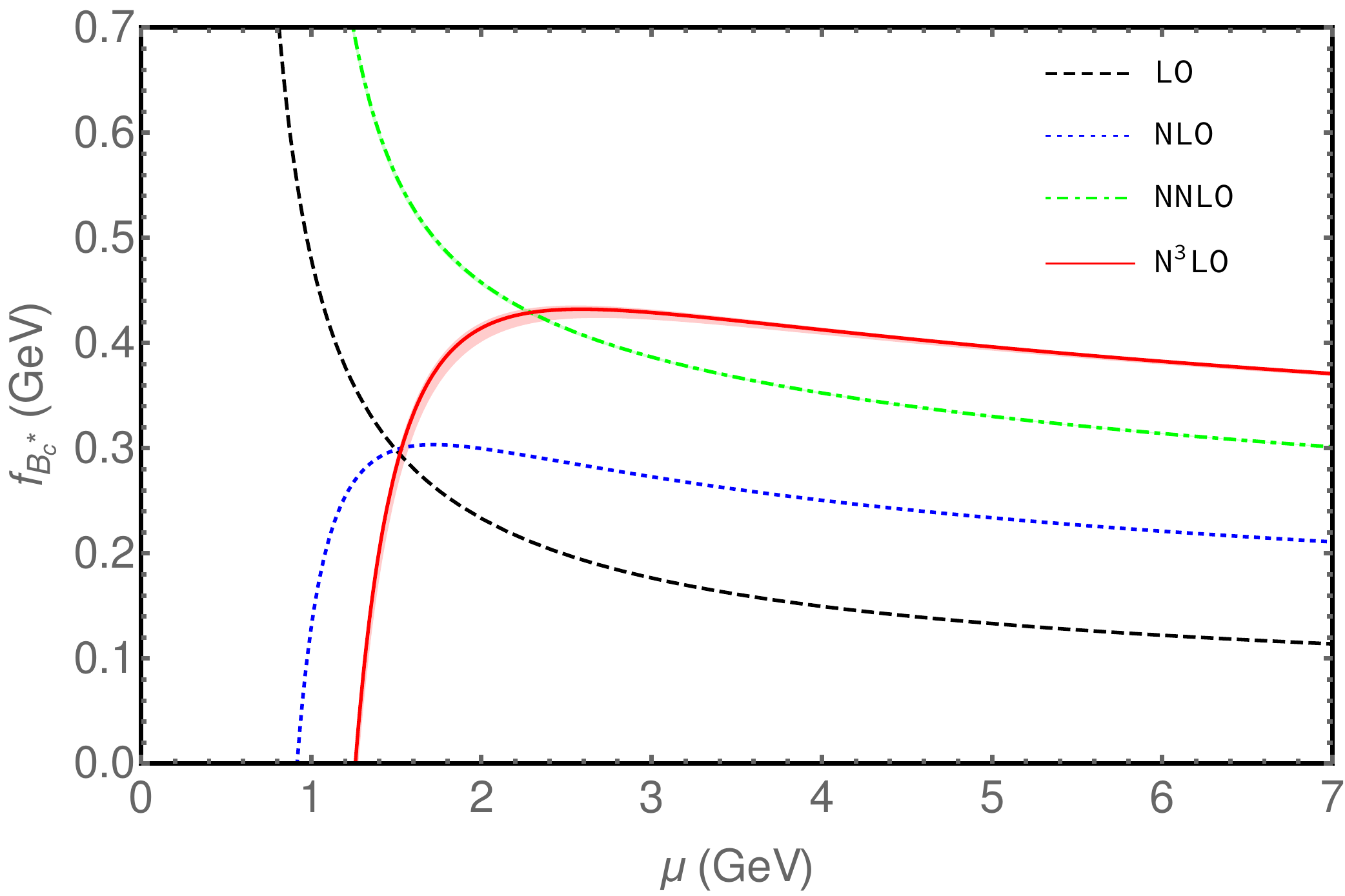}\qquad
	\includegraphics[width=0.44\textwidth]{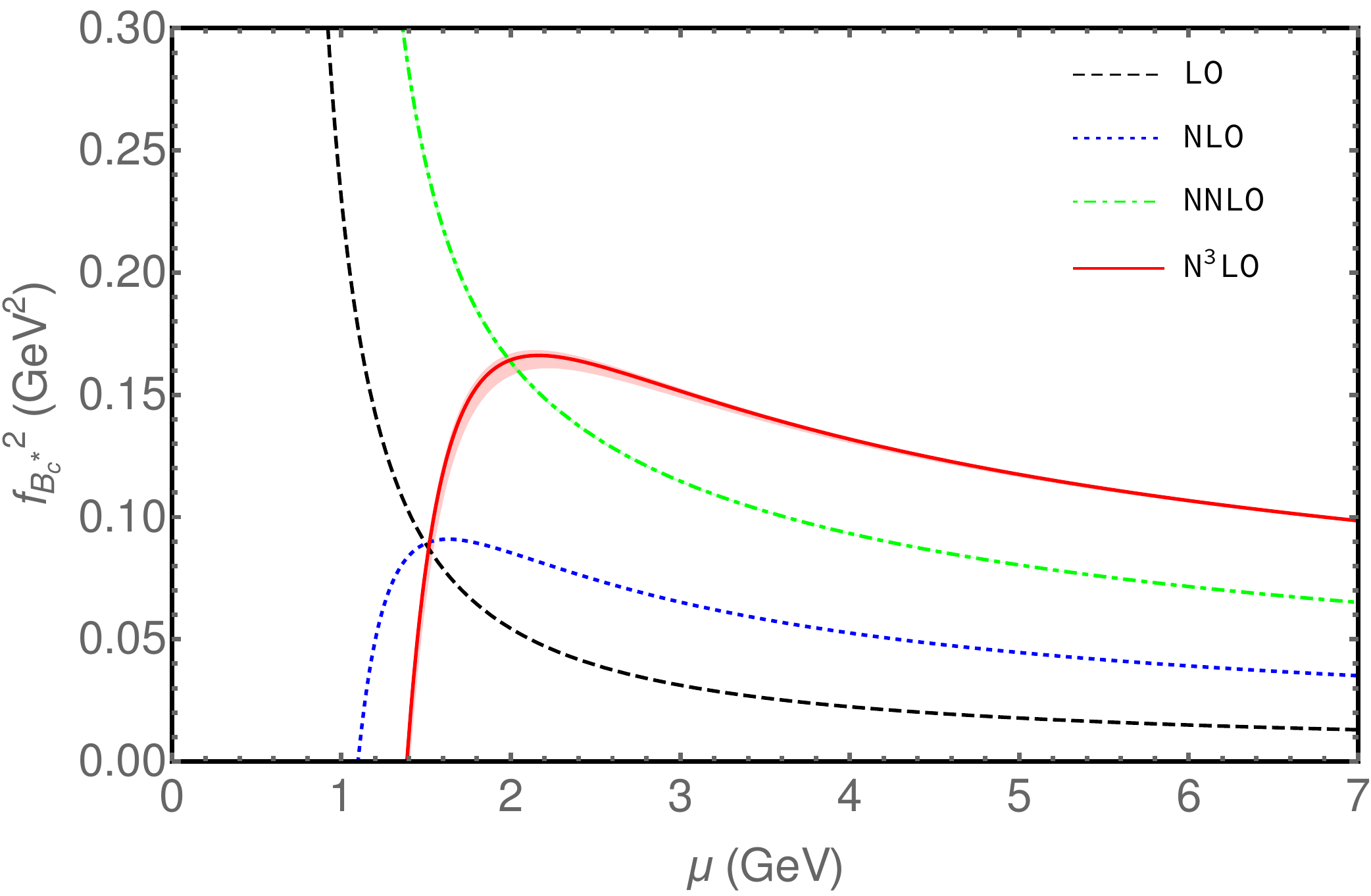}\qquad
	\includegraphics[width=0.44\textwidth]{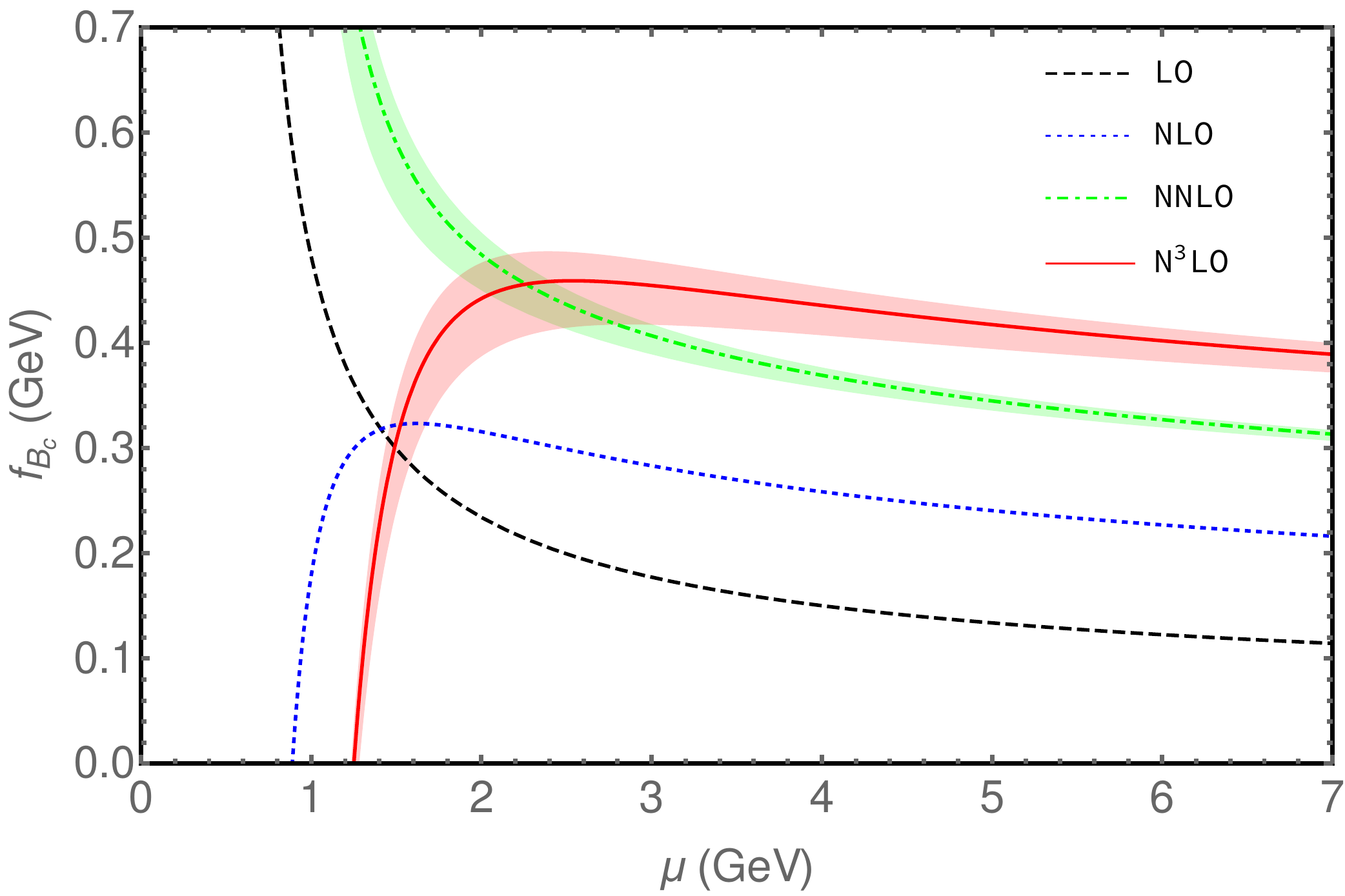}\qquad
	\includegraphics[width=0.44\textwidth]{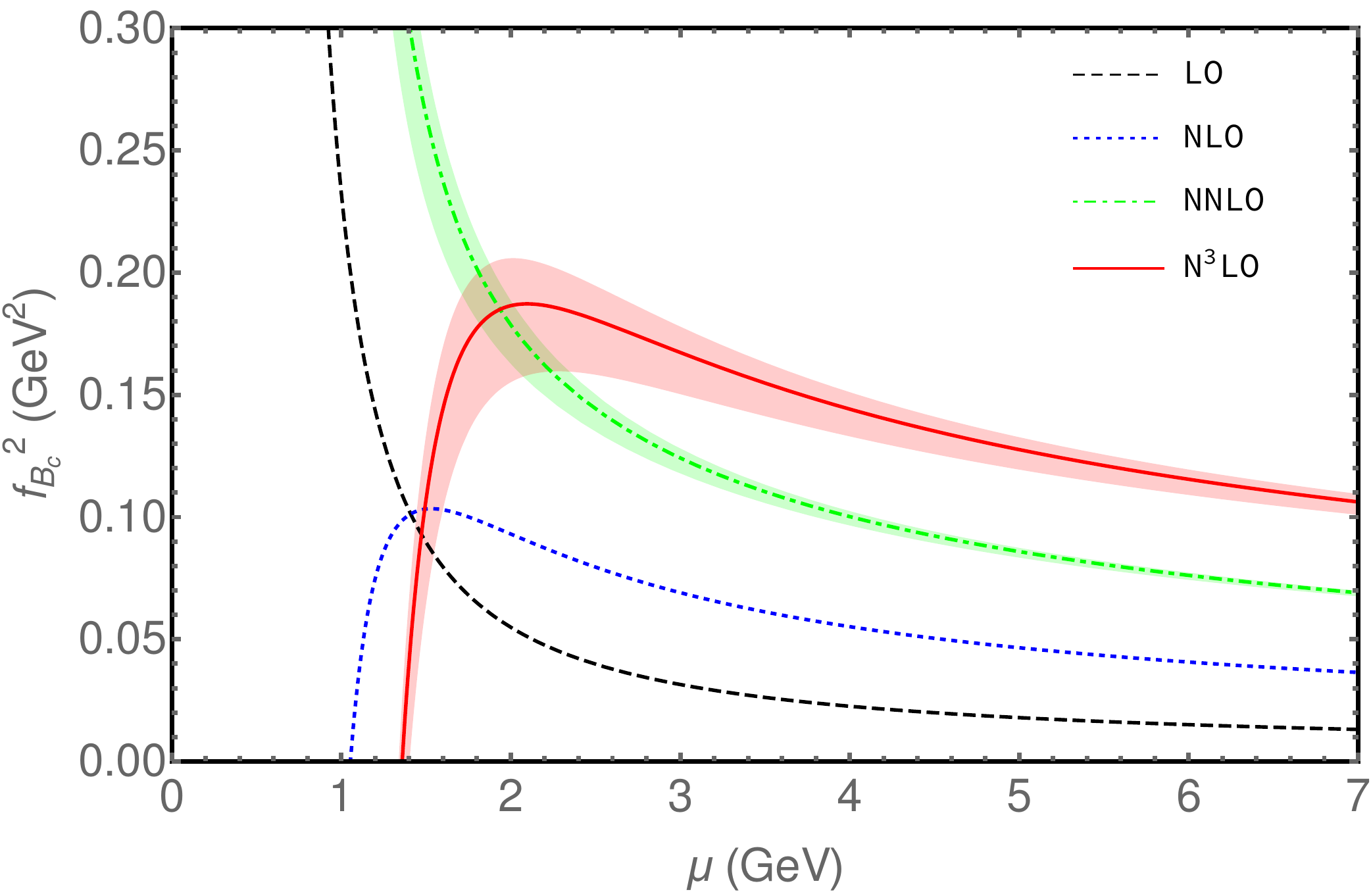}\qquad
	\caption{The renormalization scale $\mu$-dependence of the decay constants $f_{B_c^*}$, $f_{B_c}$ and the squared decay constants $f_{B_c^*}^2$, $f_{B_c}^2$  
		at LO,  NLO,  NNLO and N$^3$LO accuracy. The central values  are calculated inputting the  physical values with $y=y_c$, $\mu_f=1.2~\,\mathrm{GeV}$,  $m_b=4.75\mathrm{GeV}$ and $m_c=1.5\mathrm{GeV}$.   The error bands come from the variation of  $\mu_f$  between  7 and 0.4 $\mathrm{GeV}$. }
	\label{fig:fsmu}
\end{figure}

\begin{figure}[thb]
	\centering
	\includegraphics[width=0.44\textwidth]{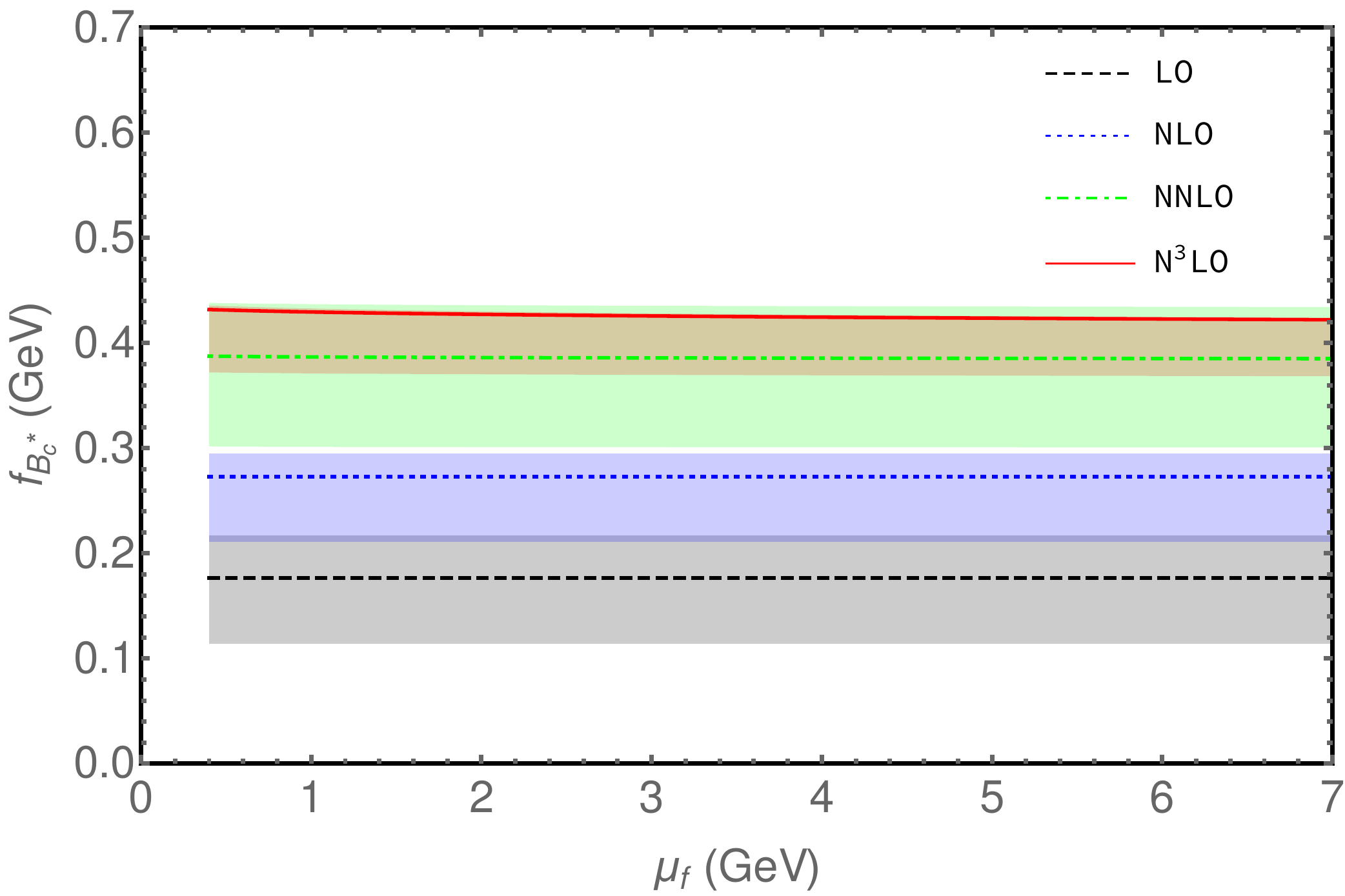}\qquad
	\includegraphics[width=0.44\textwidth]{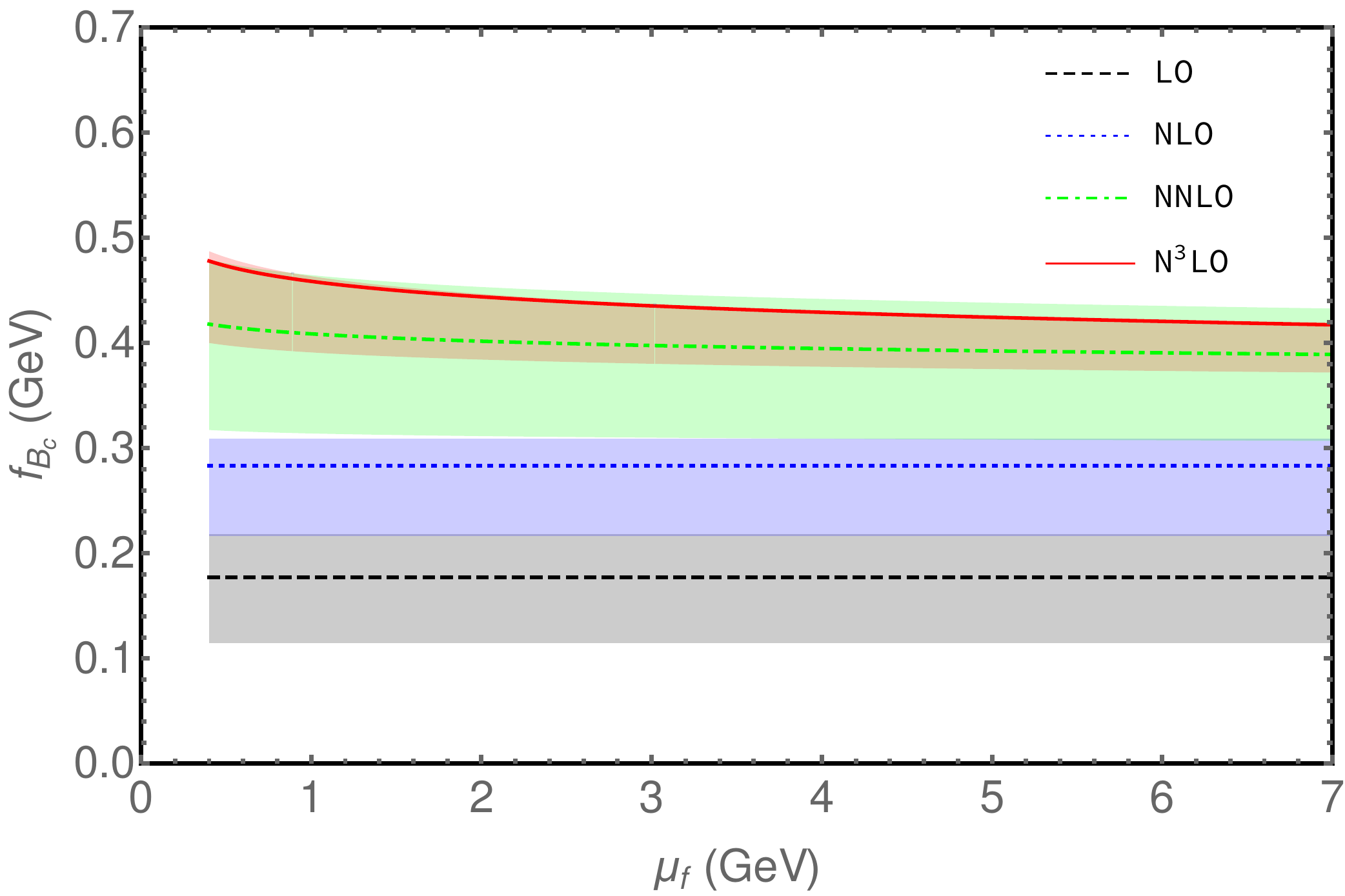}\qquad
	\caption{The factorization scale $\mu_f$-dependence of the decay constants $f_{B_c^*}$ and $f_{B_c}$ 
		at LO,  NLO,  NNLO and N$^3$LO accuracy. The central values  are calculated inputting the  physical values with $y=y_c$,  $\mu=\mu_0=3~\,\mathrm{GeV}$,  $m_b=4.75\mathrm{GeV}$ and $m_c=1.5\mathrm{GeV}$.   The error bands come from the variation of  $\mu$  between   7 and 2.2 $\mathrm{GeV}$. }
	\label{fig:fsmuf}
\end{figure}

\begin{figure}[thb]
	\centering
	\includegraphics[width=0.44\textwidth]{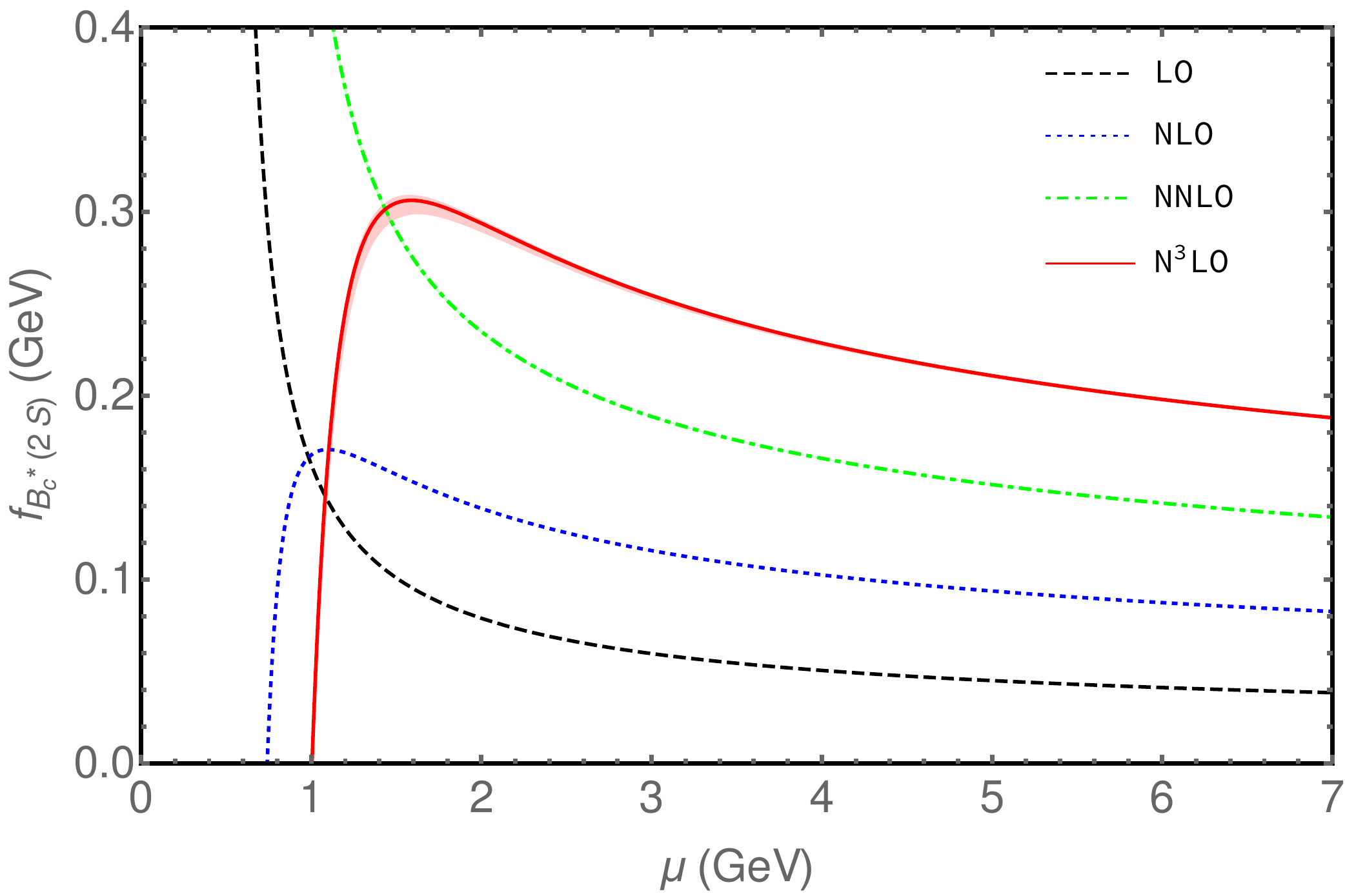}\qquad
	\includegraphics[width=0.44\textwidth]{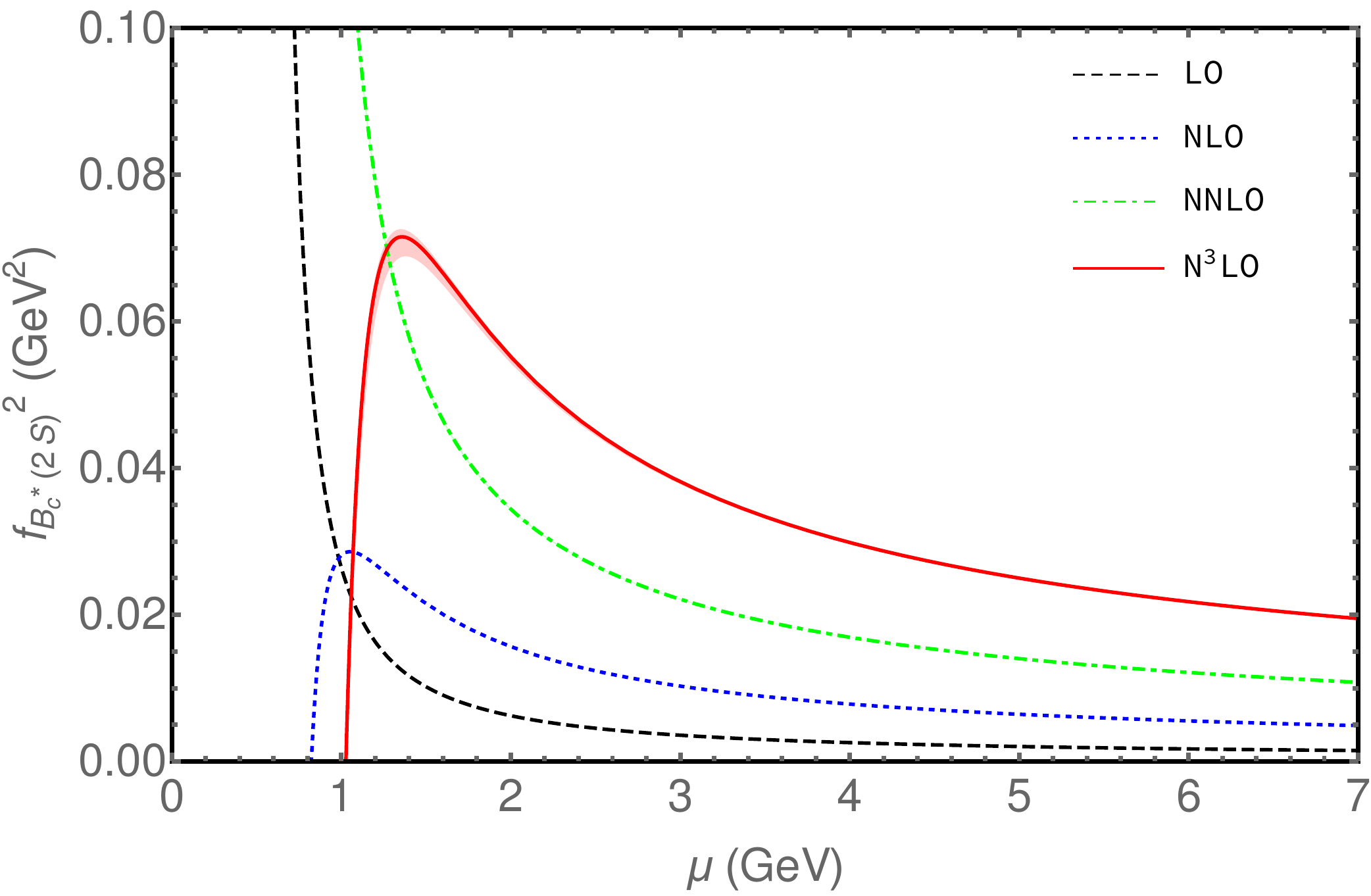}\qquad
	\includegraphics[width=0.44\textwidth]{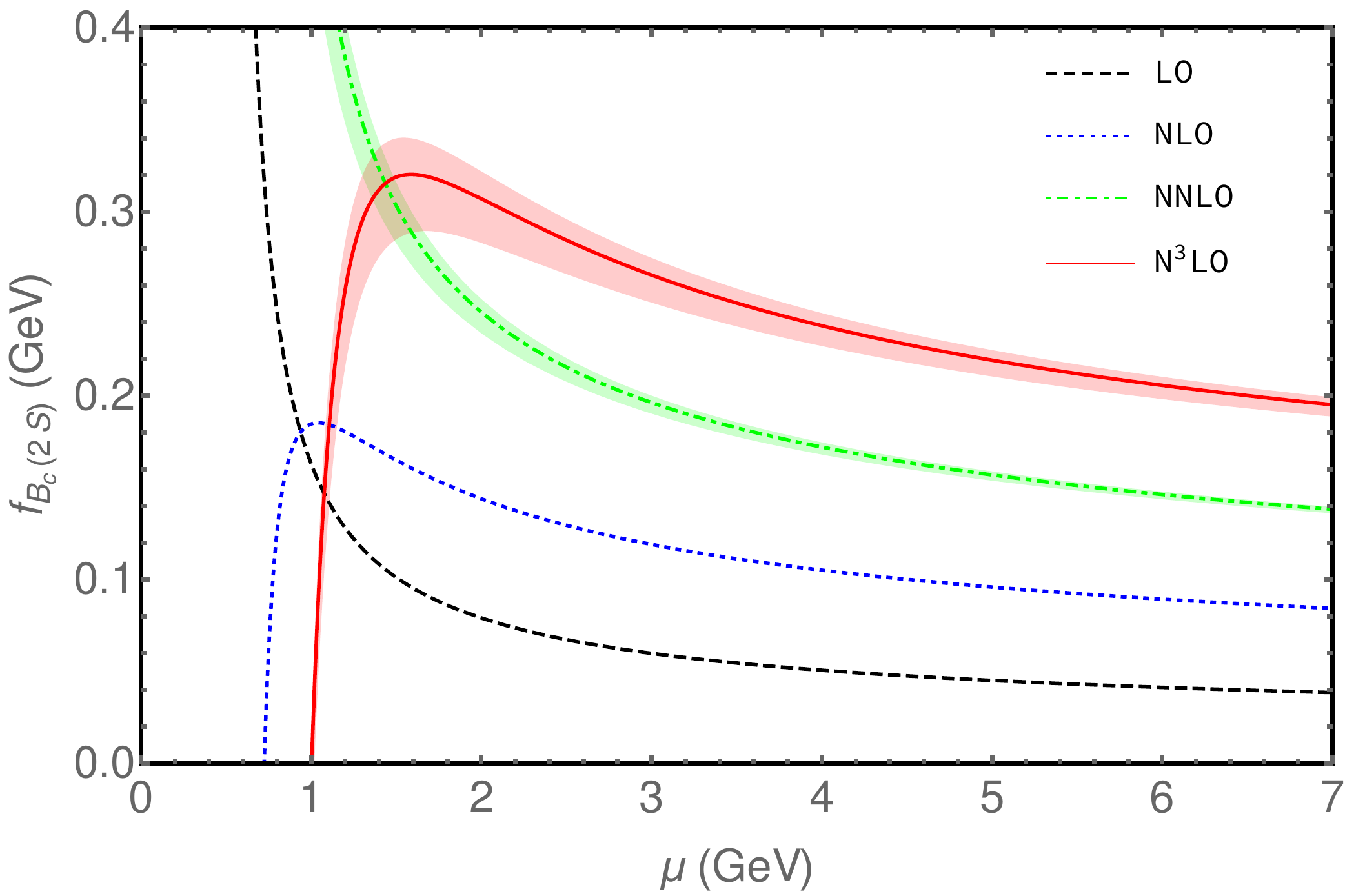}\qquad
	\includegraphics[width=0.44\textwidth]{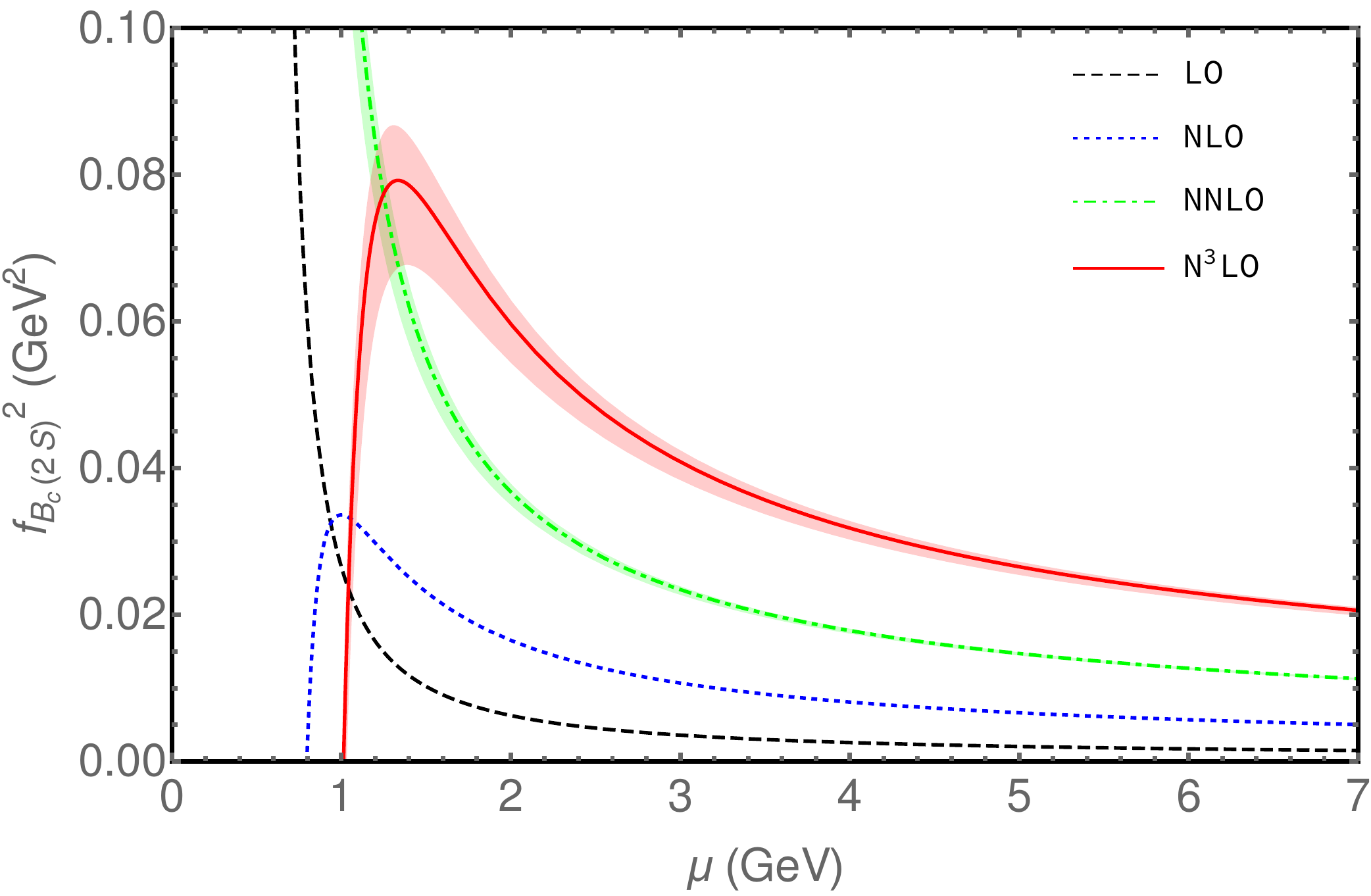}\qquad
	\caption{The same as Fig.~\ref{fig:fsmu}, but for
		  $f_{B_c^*(2S)},f_{B_c^*(2S)}^2,f_{B_c(2S)},f_{B_c(2S)}^2$, respectively. }
	\label{fig:f2smu}
\end{figure}

\begin{figure}[thb]
	\centering
	\includegraphics[width=0.44\textwidth]{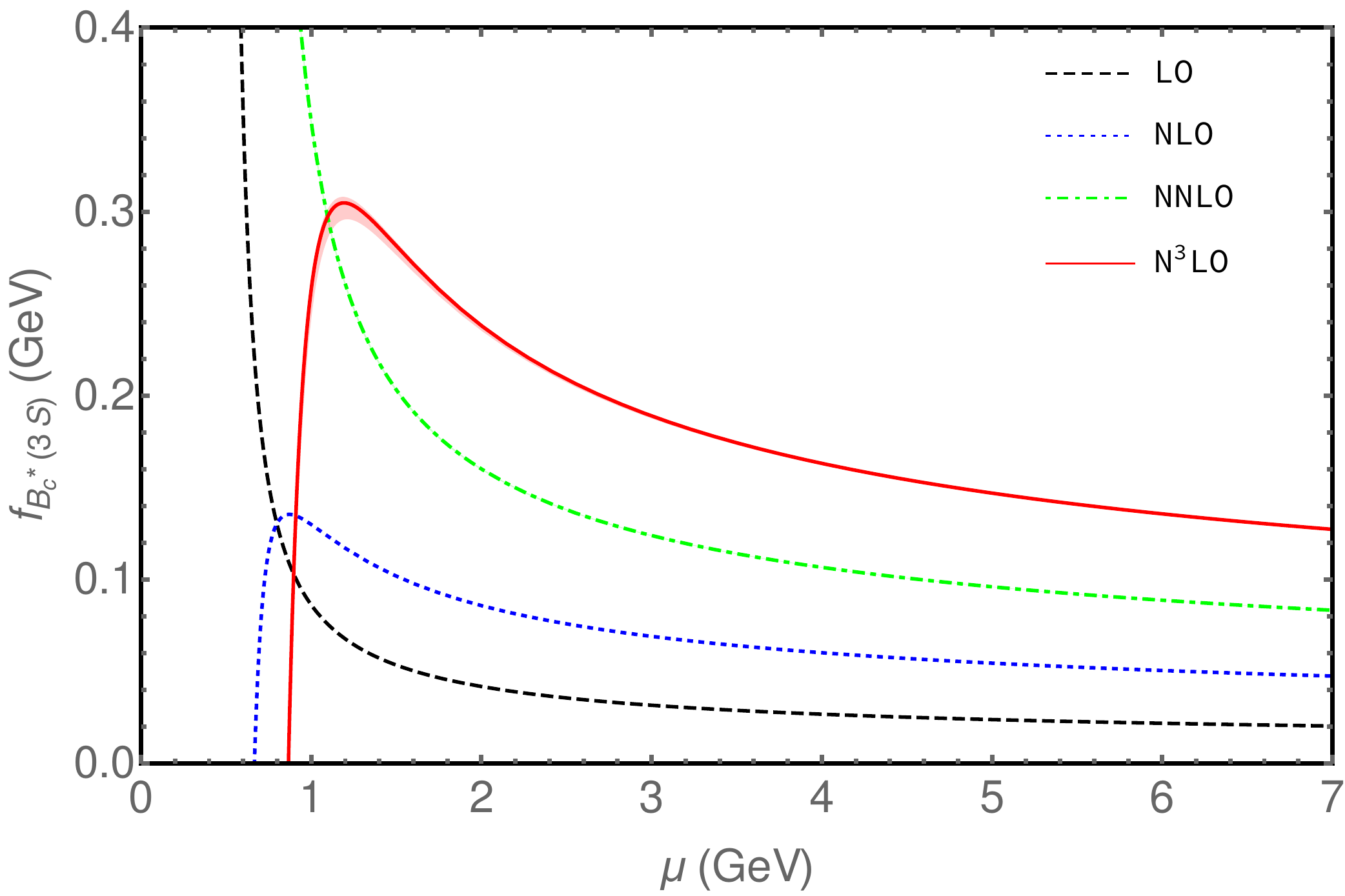}\qquad
	\includegraphics[width=0.44\textwidth]{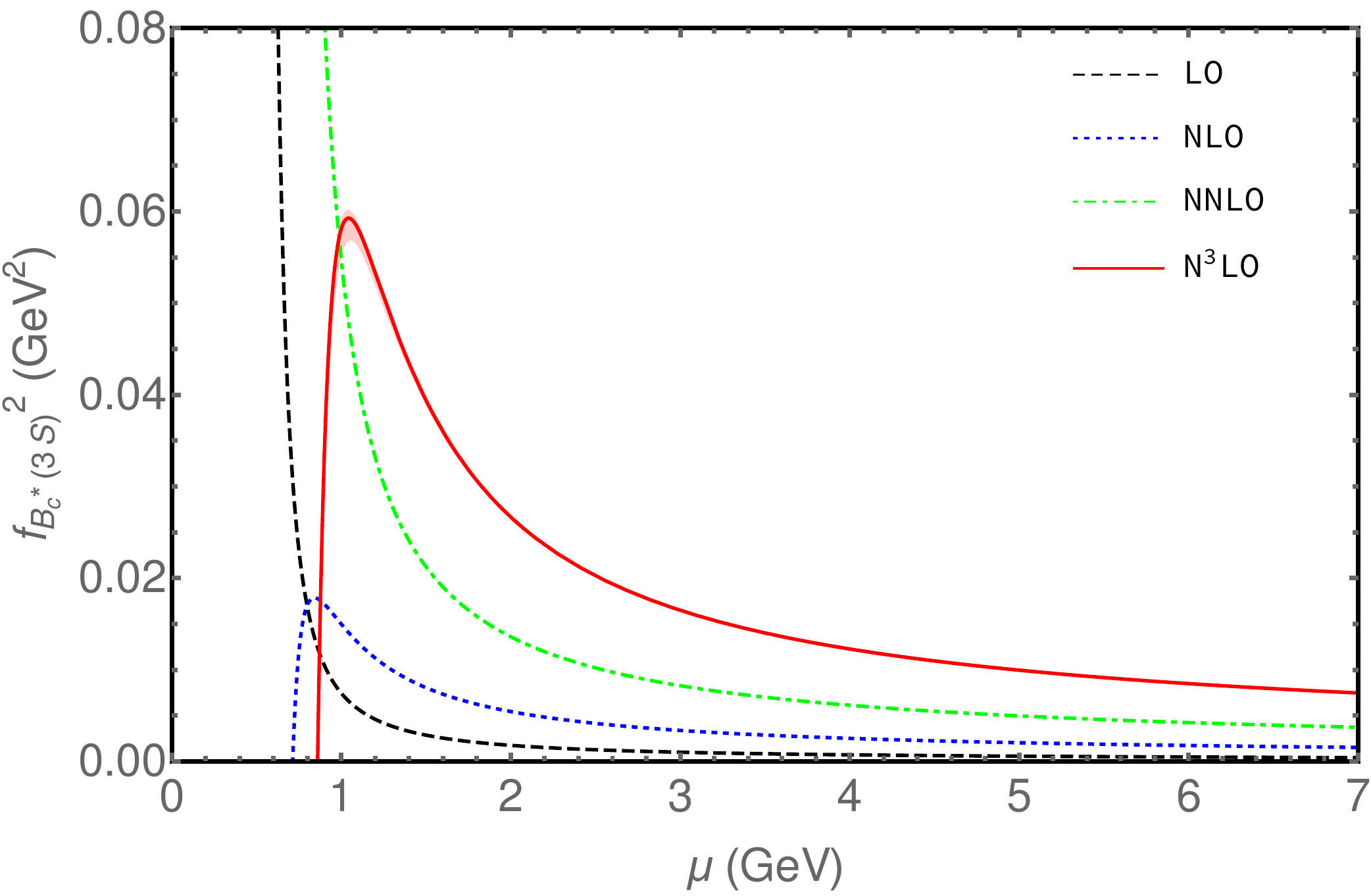}\qquad
	\includegraphics[width=0.44\textwidth]{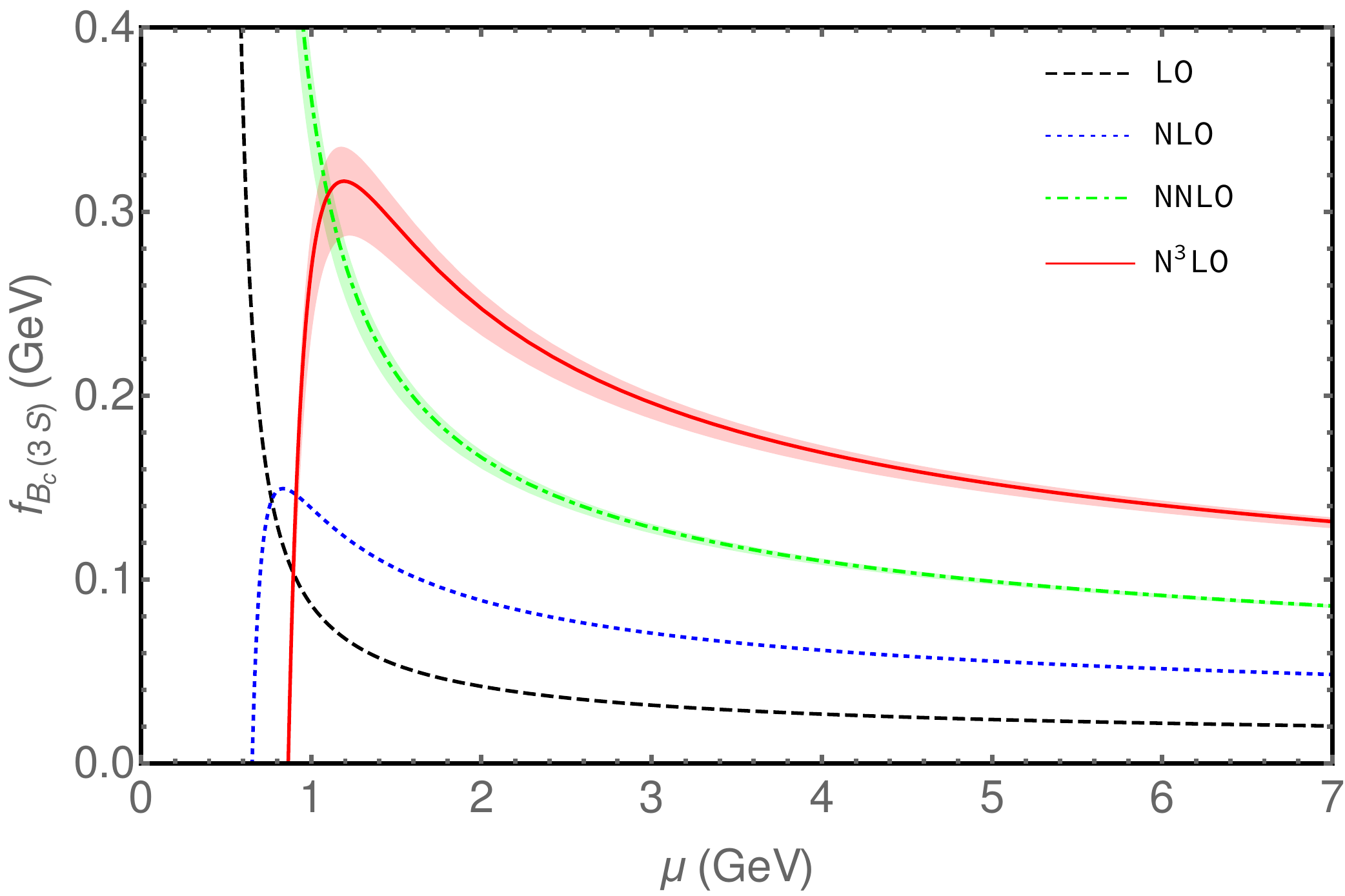}\qquad
	\includegraphics[width=0.44\textwidth]{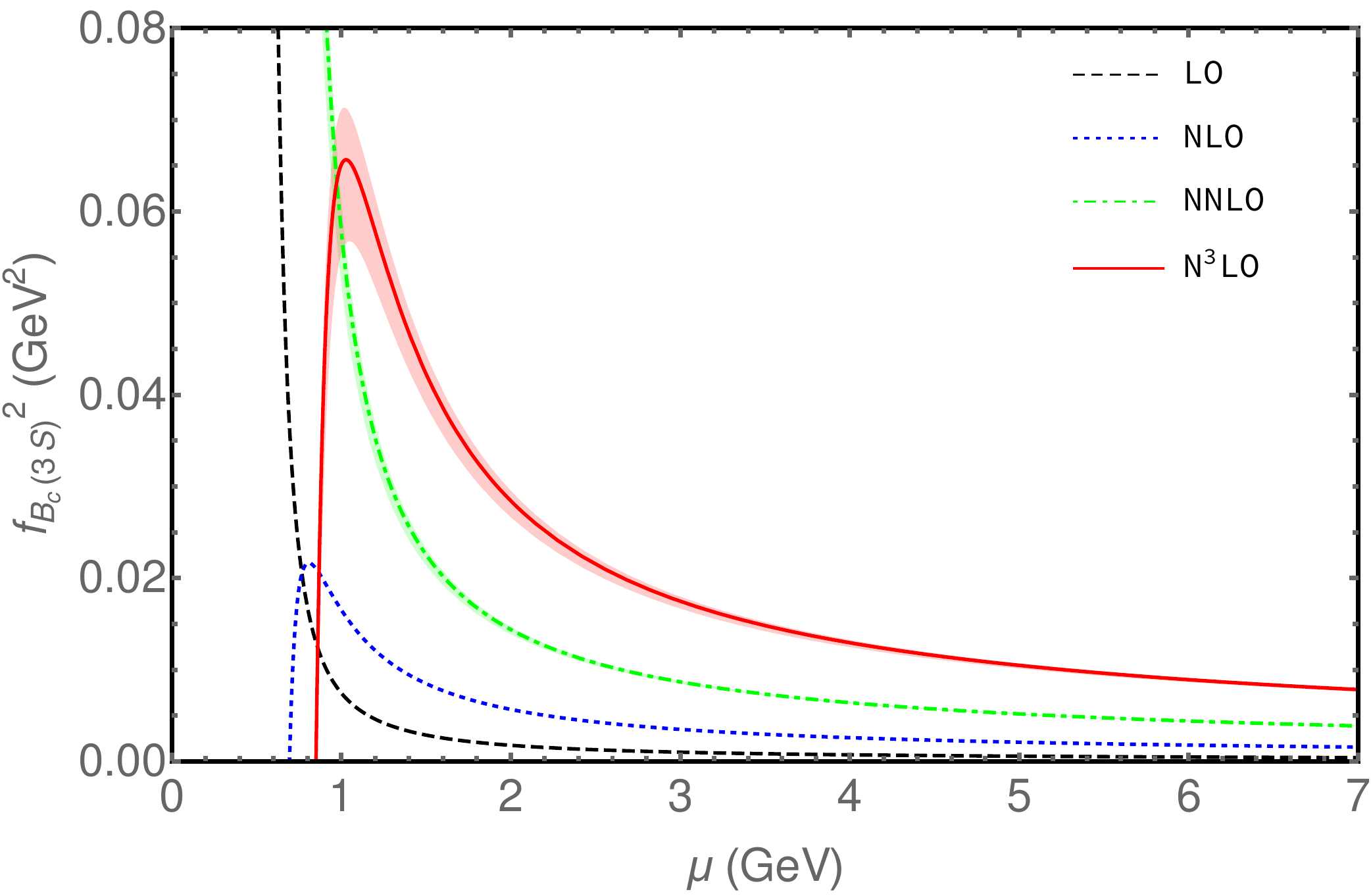}\qquad
	\caption{The same as Fig.~\ref{fig:fsmu}, but for
		$f_{B_c^*(3S)},f_{B_c^*(3S)}^2,f_{B_c(3S)},f_{B_c(3S)}^2$, respectively. }
	\label{fig:f3smu}
\end{figure}

Furthermore, we investigate the renormalization scale $\mu$-dependence  of  the  decay constants  and the squared decay constants in Fig.~\ref{fig:fsmu}, where the   middle lines, upper edges, lower  edges in the error bands correspond to  $\mu_f=1.2,~0.4,~7~\mathrm{GeV}$, respectively. Besides, we also investigate the NRQCD factorization scale  $\mu_f$-dependence of the  decay constants  in Fig.~\ref{fig:fsmuf}.
From  Figs.~(\ref{fig:fsmu}, \ref{fig:fsmuf}) one can see the following points:
\begin{enumerate}
	\item[(1)]
	Near $\mu=\mu_0=3\mathrm{GeV}$, both of the  decay constants  and the squared decay constants have a good  convergence of $\alpha_s$-expansion at N$^3$LO.
	The renormalization scale $\mu$-dependence of the decay constants and the squared decay constants has been reduced at N$^3$LO compared with NNLO.  
	The convergent decay constants imply large cancellations at $\mathcal{O}(\alpha_s^3)$ between the matching coefficients and the wave functions at the origin.

	\item[(2)]
	Comparing with  Fig.~\ref{fig:Cjmudepend},  one can  see the decay constants $f_{B_c^*}$ and the squared decay constants $f_{B_c^*}^2$ are almost free from  the factorization scale $\mu_f$, which implies the large $\mu_f$-dependence of $\mathcal{C}_{v}$
	has been canceled out by the  $\mu_f$-dependence of the wave function at the origin for the vector $B_c^*$ meson. Therefore, 
	$|\Psi_{B_c^*}(0)|$ obtained by Eq.~\eqref{bbcc2bc} is reliable.
	For the pseudoscalar $B_c$ meson, $f_{B_c}$ and $f_{B_c}^2$ still have some dependence on the the factorization scale $\mu_f$, especially in the low $\mu_f$ region and at the three-loop order, which is due to  the approximation  $|\Psi_{B_c}(0)|\approx |\Psi_{B_c^*}(0)|$ adopted in the  calculation of $f_{B_c}$ and $f_{B_c}^2$. In fact, the wave functions at the origin depend on the spin quantum numbers, which are different for $B_c$ and $B_c^*$.
	To confirm our conclusion, instead of the approximation  $|\Psi_{B_c}(0)|\approx |\Psi_{B_c^*}(0)|$, we use Eq.~\eqref{bbcc2bc} (but for the pseudoscalar version) to   obtain the NNLO result $|\Psi_{B_c}(0)|^{\rm NNLO}$ from the NNLO correction to the wave function at the origin for the lowest-lying equal-mass pseudoscalar heavy quarkonium $Q\bar{Q}$, i.e. $|\Psi_{Q\bar{Q}}^P(0)|^{\rm NNLO}$, which is available in Refs.~\cite{Penin:2004ay,Pineda:2006ri,Pineda:2011dg}. By rerunning the process, we found the factorization scale $\mu_f$-dependence of the recalculated $f_{B_c}$ and ${f_{B_c}^2}$ vanished up to NNLO.

	\item[(3)]
	Since the $\alpha_s$-expansion coefficients of the decay constants at higher-order are not small enough (see Table~\ref{tab:asseriesallv} and Table~\ref{tab:asseriesallp}),
	the  convergence of $\alpha_s$-expansion for the squared decay constants is not as good as the decay constants.
	After squaring the decay constants,  the N$^3$LO results show larger $\mu$-dependence compared with LO though both the decay constants and the squared decay constants are renormalization-group running invariant as Eq.~\eqref{runinvariance}. The origin of the problems may be related to the nonperturbative effect~\cite{Rauh:2018vsv}. In addition, the NRQCD and pNRQCD effective theory require the assumption $m_r u^2\gg\Lambda_{QCD}$, which is still questionable for the $c\bar{b}$    meson system~\cite{Egner:2021lxd,Beneke:2014qea,Soto:2004bb}.

	\item[(4)]
	To further study the perturbative and nonperturbative nature of the $c\bar{b}$ system, 
	we apply our calculation procedure to the (squared) decay constants of the  excited states: $B_c^*(2S),B_c(2S),B_c^*(3S),B_c(3S)$ with masses $6.900,6.871,7.260,7.240\mathrm{GeV}$~\cite{Workman:2022ynf,Chaturvedi:2022pmn}, respectively.
	By Eq.~\eqref{fBcv} and Eq.~\eqref{fBc} with the above mesons masses,  the binding energies~\cite{Peset:2015vvi} and the wave functions at the origin~\cite{Schuller:2008rxa} for the  excited states, we calculate the (squared) decay constants
	 up to N$^3$LO and plot them in Fig.~\ref{fig:f2smu} and Fig.~\ref{fig:f3smu}. Comparing the figures with Fig.~\ref{fig:fsmu}, 
  we see	the perturbative expansions of the (squared) decay constants for the  excited states are not convergent and show larger renormalization scale  dependence than the lowest-lying ground states $B_c^*(1S)$ and $B_c(1S)$, which suggests the nonperturbative contribution is  non-negligible and the excited states may appear to be in the nonperturbative regime~\cite{Soto:2004bb,Beneke:2005hg}.
	 Our conclusion is that, for the $c\bar{b}$ system, 
	 the	 perturbative computation by employing the NRQCD and pNRQCD 
	  is indeed applicable  to the lowest-lying  ground states $B_c^*(1S),B_c(1S)$.
	  But its  applicability for  the excited states  can not be guarantied at present, more studies are clearly required.
	
\end{enumerate}

With $f_{B_c^*}^2$ and $f_{B_c}^2$ in hand, the leptonic decay widths  of     the vector  $B_c^*$ meson and the pseudoscalar  $B_c$ meson    can be straightforwardly evaluated by the following formulas~\cite{Zhou:2017svh,Yang:2021crs}:
\begin{align}
	\Gamma(B_c^{*+} \to l^{+}  {\nu}_{l})&=\frac{{|V_{cb}|}^2}{12\pi}  G_F^2 m_{B_c^*}^3 \left(1-\frac{m_l^2}{m_{B_c^*}^2}\right)^2 \left(1+\frac{m_l^2}{2 m_{B_c^*}^2}\right) {f_{B_c^*}^2}
	\\
	\Gamma(B^+_c \to l^{+}  {\nu}_{l})&=\frac{{|V_{cb}|}^2}{8\pi}  G_F^2 m_{B_c} m_l^2 \left(1-\frac{m_l^2}{m_{B_c}^2}\right)^2 {f_{B_c}^2}.
\end{align}
In the numerical calculations, the  values of the following  input parameters have been used implicitly unless otherwise stated~\cite{Workman:2022ynf,Mathur:2018epb,Gomez-Rocha:2016cji,Zhou:2017svh,Yang:2021crs,Gregory:2009hq,Fulcher:1998ka,Ikhdair:2003ry,Martin-Gonzalez:2022qwd}:				
\begin{align}
	&	m_e=0.510999 \mathrm{MeV}, ~m_{\mu}=0.10566 \mathrm{GeV}, ~m_{\tau}=1.777 	\mathrm{GeV},
	\nonumber\\&
	V_{cb}=0.0408, ~ G_F=1.16638 \times 10^{-5} \mathrm{GeV}^{-2},
	~	\tau_{B_c}=0.51  \mathrm{ps}, ~	\Gamma_{\mathrm{tot}}(B_c^*)=60  \mathrm{eV}.
\end{align}

In Tables \ref{tab:fsnum}-\ref{tab:brsnum},  we present the perturbative QCD predictions up to N$^3$LO for the decay constants $f_{B^*_c}$ and $f_{B_c}$,
the leptonic decay widths  $\Gamma(B_c^{*+} \to l^{+}  {\nu}_{l})$ and $\Gamma (B_c^+ \to l^{+}  {\nu}_{l})$,  as well as the corresponding branching ratios
 ${\cal B} (B_c^{*+} \to l^{+}  {\nu}_{l} )$ and ${\cal B} (B_c^+ \to l^{+}  {\nu}_{l})$   with $l=(e,\mu,\tau)$.		
 From the numeric results, one can see the major uncertainties come from the error of the renormalization scale $\mu$, the uncertainties from the scale power $y$ are about half the magnitude of the uncertainties from $\mu$,  and 	the influence of the factorization scale $\mu_f$	is the smallest. One can also find the N$^3$LO corrections have reduced the uncertainties from $\mu$ compared with the NNLO corrections.
	
\begin{table}[thb]\footnotesize
	\begin{center}
		\caption{The theoretical predictions  of the  decay constants  $f_{B_c^*}$ and $f_{B_c}$  at the LO, NLO, NNLO and N$^3$LO level.
						The central values  are calculated by using the  physical values with $y=y_c$, $\mu_f=1.2~\mathrm{GeV}$, $\mu=\mu_0=3\mathrm{GeV}$, $m_b=4.75\mathrm{GeV}$ and $m_c=1.5\mathrm{GeV}$.
			The errors are estimated by varying $y$ from 0.4 to 0.3, $\mu_f$  from   7 to 0.4 $\mathrm{GeV}$, $\mu$  from   7 to 2.2 $\mathrm{GeV}$, respectively.}
		\label{tab:fsnum}
		\renewcommand\arraystretch{2}
		\tabcolsep=0.3cm
		\begin{tabular}{ c c c c c}
			\hline\hline
			& LO         &  NLO                   & NNLO     & N$^3$LO
			\\  \hline
			$f_{B_c^*}(\mathrm{10^{-1}GeV})$	 & $1.77^{+0.12-0-0.63}_{-0.18+0+0.41}$ & $2.73^{+0.14-0-0.62}_{-0.22+0+0.22}$   &    $3.87^{+0.17-0.01-0.85}_{-0.27+0.01+0.50}$  &     $4.29^{+0.11-0.07-0.58}_{-0.19+0.03+0.01}$
			\\  \hline
			$f_{B_c}(\mathrm{10^{-1}GeV})$	 & $1.77^{+0.12-0-0.63}_{-0.18+0+0.41}$ &     $2.83^{+0.14-0-0.67}_{-0.23+0+0.26}$
			 & $4.07^{+0.18-0.18-0.94}_{-0.29+0.11+0.55}$   &    $4.55^{+0.12-0.37-0.66}_{-0.21+0.23+0.01}$ 	
			\\		\hline \hline
		\end{tabular}
	\end{center}
\end{table}

\begin{table}[thb]\scriptsize
	\begin{center}
		\caption{The same as Table ~\ref{tab:fsnum}, but for   $\Gamma( B_c^{*+}/B_c^{+} \to l^{+}  {\nu}_{l})$ with $l=(e,\mu,\tau)$.}
		\label{tab:gamsnum}
		\renewcommand\arraystretch{2}
		\tabcolsep=0.15cm
		\begin{tabular}{ c c c c c}
			\hline\hline
			& LO         &  NLO                   & NNLO     & N$^3$LO
			\\  \hline
			$\Gamma(B^{*+}_c \to e^{+}  {\nu}_{e}) ( 10^{-13} {\rm GeV})$	 & $0.48^{+0.06-0-0.28}_{-0.09+0+0.24}$ & $0.99^{+0.11-0-0.46}_{-0.17+0+0.24}$   &    $1.75^{+0.17-0.01-0.76}_{-0.26+0.01+0.52}$  &     $2.31^{+0.17-0.04-0.81}_{-0.28+0.02+0.22}$
				\\  \hline
			$\Gamma(B^{*+}_c \to {\mu}^{+}  {\nu}_{\mu}) ( 10^{-13} {\rm GeV})$	 & $0.48^{+0.06-0-0.28}_{-0.09+0+0.24}$ & $0.99^{+0.11-0-0.46}_{-0.17+0+0.24}$   &    $1.75^{+0.17-0.01-0.76}_{-0.26+0.01+0.52}$  &     $2.31^{+0.17-0.04-0.81}_{-0.28+0.02+0.22}$
				\\  \hline
			$\Gamma(B^{*+}_c \to {\tau}^{+}  {\nu}_{\tau}) ( 10^{-13} {\rm GeV})$	 & $0.42^{+0.06-0-0.24}_{-0.08+0+0.22}$ & $0.88^{+0.10-0-0.40}_{-0.15+0+0.21}$  &     $1.54^{+0.15-0.01-0.67}_{-0.23+0.00+0.46}$	  &    $2.04^{+0.15-0.04-0.71}_{-0.24+0.02+0.20}$ 		
			\\  \hline
			$\Gamma(B^{+}_c \to e^{+}  {\nu}_{e}) ( 10^{-21} {\rm GeV})$	 & $0.46^{+0.06-0-0.27}_{-0.09+0+0.24}$ & $1.02^{+0.11-0-0.48}_{-0.17+0+0.27}$   &    $1.83^{+0.18-0.09-0.81}_{-0.28+0.06+0.56}$  &     $2.47^{+0.19-0.25-0.90}_{-0.30+0.16+0.28}$	
			\\  \hline
				$\Gamma(B^{+}_c \to {\mu}^{+}  {\nu}_{\mu}) ( 10^{-16} {\rm GeV})$	 & $0.20^{+0.03-0-0.12}_{-0.04+0+0.10}$ & $0.44^{+0.05-0-0.21}_{-0.07+0+0.12}$   &    $0.78^{+0.08-0.04-0.35}_{-0.12+0.02+0.24}$  &     $1.06^{+0.08-0.11-0.39}_{-0.13+0.07+0.12}$	
			\\  \hline
				$\Gamma(B^{+}_c \to {\tau}^{+}  {\nu}_{\tau}) ( 10^{-14} {\rm GeV})$	 & $0.48^{+0.06-0-0.28}_{-0.09+0+0.24}$ & $1.04^{+0.12-0-0.49}_{-0.17+0+0.28}$   &    $1.87^{+0.19-0.10-0.83}_{-0.29+0.06+0.57}$  &     $2.53^{+0.19-0.26-0.92}_{-0.31+0.16+0.29}$				
			\\		\hline \hline
		\end{tabular}
	\end{center}
\end{table}

\begin{table}[thb]\scriptsize
	\begin{center}
		\caption{The same as Table ~\ref{tab:fsnum}, but for   ${\cal B} ( B_c^{*+}/B_c^{+} \to l^{+}  {\nu}_{l})$ with $l=(e,\mu,\tau)$.}
		\label{tab:brsnum}
		\renewcommand\arraystretch{2}
		\tabcolsep=0.15cm
		\begin{tabular}{ c c c c c}
			\hline\hline
			& LO         &  NLO                   & NNLO     & N$^3$LO
			\\  \hline
			${\cal B} (B^{*+}_c \to e^{+}  {\nu}_{e}) (\times 10^{-6} )$	 & $0.79^{+0.11-0-0.46}_{-0.16+0+0.41}$ & $1.66^{+0.18-0-0.76}_{-0.28+0+0.40}$   &    $2.91^{+0.29-0.01-1.26}_{-0.44+0.01+0.86}$  &     $3.85^{+0.29-0.07-1.35}_{-0.46+0.03+0.37}$
			\\  \hline
			${\cal B} (B^{*+}_c \to {\mu}^{+}  {\nu}_{\mu}) (\times 10^{-6} )$	  & $0.79^{+0.11-0-0.46}_{-0.16+0+0.41}$ & $1.66^{+0.18-0-0.76}_{-0.28+0+0.40}$   &    $2.91^{+0.29-0.01-1.26}_{-0.44+0.01+0.86}$  &     $3.85^{+0.29-0.07-1.35}_{-0.46+0.03+0.37}$  
			\\  \hline
			${\cal B} (B^{*+}_c \to {\tau}^{+}  {\nu}_{\tau}) (\times 10^{-6} )$	  & $0.70^{+0.09-0-0.41}_{-0.14+0+0.36}$ & $1.46^{+0.16-0-0.67}_{-0.24+0+0.35}$   &    $2.57^{+0.25-0.01-1.11}_{-0.39+0.01+0.76}$  &     $3.40^{+0.25-0.06-1.19}_{-0.41+0.03+0.33}$
			\\  \hline
			${\cal B} (B^{+}_c \to e^{+}  {\nu}_{e}) (\times 10^{-9} )$	  & $0.36^{+0.05-0-0.21}_{-0.07+0+0.18}$ & $0.79^{+0.09-0-0.37}_{-0.13+0+0.21}$   &    $1.42^{+0.14-0.07-0.63}_{-0.22+0.04+0.43}$  &     $1.91^{+0.15-0.19-0.70}_{-0.23+0.12+0.22}$
			\\  \hline
			${\cal B} (B^{+}_c \to {\mu}^{+}  {\nu}_{\mu}) (\times 10^{-5} )$	  & $1.54^{+0.21-0-0.90}_{-0.30+0+0.79}$ & $3.37^{+0.38-0-1.59}_{-0.57+0+0.89}$   &    $6.07^{+0.61-0.31-2.69}_{-0.92+0.19+1.86}$  &     $8.18^{+0.63-0.83-2.99}_{-1.00+0.52+0.94}$
			\\  \hline
			${\cal B} (B^{+}_c \to {\tau}^{+}  {\nu}_{\tau}) (\times 10^{-2} )$	  & $0.37^{+0.05-0-0.21}_{-0.07+0+0.19}$ & $0.81^{+0.09-0-0.38}_{-0.14+0+0.21}$   &    $1.45^{+0.15-0.07-0.64}_{-0.22+0.05+0.44}$  &     $1.96^{+0.15-0.20-0.72}_{-0.24+0.12+0.23}$
			\\		\hline \hline
		\end{tabular}
	\end{center}
\end{table}

\section{Summary~\label{Summary}}

In this paper, we have performed  the N$^3$LO QCD corrections to the matching coefficients for the heavy flavor-changing   currents allowing for  the  vector, axial-vector, scalar		and pseudo-scalar   cases   within the framework of NRQCD factorization.
We have obtained the analytical expressions of the three-loop renormalization constants and corresponding three-loop anomalous dimensions of the all four NRQCD heavy flavor-changing   currents.
Meanwhile, the three-loop matching coefficients have also been obtained with high numerical accuracy, which are   building blocks of   physical quantities  and   are also helpful to analyze the threshold behaviours when the heavy bottom and charm quarks are combined into the $c\bar{b}$ mesons.
The obtained N$^3$LO QCD  corrections to  the matching coefficients  are considerably large compared with lower-order corrections, and  exhibit
stronger dependence on the QCD renormalization scale and NRQCD factorization scale at higher order.

For a  reliable  prediction  of the physical  observables    within the NRQCD and pNRQCD  effective theory, we employ the scale relation to obtain the N$^3$LO corrections to the wave functions at the origin for $B_c$ and $B_c^*$ from the known results in equal-mass  pNRQCD.  Then we  combine the  higher   order  QCD  corrections to  the matching coefficients       with contributions from   higher   order  corrections  to the wave functions at the origin,  the binding energies and the matching coefficients at $\mathcal{O}(u^2)$.
  We find a large cancellation at the third order QCD correction between    the   matching coefficient and the wave function at the origin. The resultant complete N$^3$LO corrections to the decay constants $f_{B_c}$ and $f_{B_c^*}$ can become convergent and have small dependence on  the  NRQCD factorization scale $\mu_f$  and  the  QCD renormalization scale $\mu$, as illustrated in Figs.~(\ref{fig:fsmu}, \ref{fig:fsmuf}).     
  
In Tables~\ref{tab:fsnum}-\ref{tab:brsnum},  we present the perturbative   QCD predictions for the numerical values of the    decay constants $f_{B_c}$ and $f_{B_c^*}$, 
the leptonic decay widths  ${\Gamma} ( B_c^{+}/B_c^{*+} \to l^{+}  {\nu}_{l})$   and the corresponding branching ratios 
 ${\cal B} ( B_c^{+}/B_c^{*+} \to l^{+}  {\nu}_{l})$ with $l=(e,\mu,\tau)$   respectively.
All these predictions     will     be tested in the on-going and future precision flavor physics experiments.

\hspace{2cm}

\noindent {\bf Acknowledgments:} We thank   J. H. Piclum and  A. Onishchenko  for many helpful discussions.  This work is supported by NSFC under grant No.~11775117 and No.~12075124,  and by Natural Science Foundation of Jiangsu under Grant No.~BK20211267.

\hspace{2cm}

\bibliographystyle{JHEP}
\bibliography{refs}

\end{document}